\documentclass[aps,onecolumn,11pt,floatfix,altaffilletter,
               tightenlines,showpacs,showkeys,notitlepage,preprintnumbers,nofootinbib]{revtex4-1}
\pdfoutput=1
\usepackage[colorlinks=true,citecolor=blue,linkcolor=blue]{hyperref}
\usepackage[normalem]{ulem}
\usepackage{amsmath,amssymb}
\usepackage{mathtools}
\usepackage{epsfig}
\usepackage{graphicx}               
\usepackage{url}
\usepackage{color}
\usepackage{multirow}
\usepackage{placeins}
\usepackage[dvipsnames]{xcolor}
\usepackage{epstopdf}
\usepackage{slashed} 

\usepackage[capitalise]{cleveref}
 \usepackage{gensymb}
 
\allowdisplaybreaks

\bibpunct{[}{]}{,}{n}{}{,}

\renewcommand{\vec}[1]{{\mathbf{#1}}}


\begin{document}
\preprint{MPP-2020-28}
\title{Revisiting Neutrino Self-Interaction Constraints from $Z$ and $\tau$ decays}

\author{Vedran Brdar$^{1}$}             \email{vbrdar@mpi-hd.mpg.de}
\author{Manfred Lindner$^{1}$}	       \email{lindner@mpi-hd.mpg.de }
\author{Stefan Vogl$^{2}$}            \email{stefan.vogl@mpp.mpg.de}
\author{Xun-Jie Xu$^{1}$\,}            \email{xunjie.xu@mpi-hd.mpg.de}  
\affiliation{$^1$Max-Planck-Institut  f{\"u}r  Kernphysik,  69117  Heidelberg,  Germany\\
$^2$Max-Planck-Institut  f{\"u}r Physik, 80805 M\"unchen, Germany   
 }

\begin{abstract}
\noindent
Given the elusive nature of neutrinos, their self-interaction is particularly difficult to probe. Nevertheless, upper limits on the strength of such an interaction can be set by using data from terrestrial experiments. In this work we focus on additional contributions to the invisible decay width of $Z$ boson as well as the leptonic $\tau$ decay width in the presence of a neutrino coupling to a relatively light scalar.     
For invisible $Z$ decays we derive a complete set of constraints by considering both three-body bremsstrahlung as well as the loop correction to two-body decays. While the latter is usually regarded to give rather weak limits we find that through the interference with the Standard Model diagram it actually yields a competitive constraint. As far as leptonic decays of $\tau$ are concerned, we derive a limit on neutrino self-interactions that is valid across the whole mass range of a light scalar mediator. Our bounds on the neutrino self-interaction are leading for $m_\phi \gtrsim 300$ MeV and interactions that prefer $\nu_\tau$. Bounds on such $\nu$-philic scalar are particularly relevant in light of the recently proposed alleviation of the Hubble tension in the presence of such couplings.  

\end{abstract}

\maketitle

\section{Introduction}
\noindent
Recent studies have revealed a discrepancy between local measurements of the Hubble constant~\cite{Riess:2016jrr,Riess:2019cxk,Wong:2019kwg} and those obtained by analyzing the Cosmic Microwave Background (CMB) data~\cite{Aghanim:2018eyx} at a $\gtrsim 4\sigma$ level. This has sparked an ongoing controversy in cosmology and the search for potential solutions is currently ongoing. 
At the moment the origin of the Hubble tension is unclear; potential solutions include, for example,
early dark energy \cite{Poulin:2018cxd}, light dark matter \cite{Alcaniz:2019kah}, majorons \cite{Escudero:2019gvw},
dark matter neutrino interactions \cite{DiValentino:2017oaw}, certain classes of non-Gaussian primordial fluctuations \cite{Adhikari:2019fvb} or, more prosaically, underestimated systematics \cite{Shanks:2018rka}. Most of these ideas fall clearly into the realm of cosmology and astrophysics and cannot be tested in laboratory experiments. 
However, it was proposed recently that strong neutrino self-interactions~($\nu$SI) can alleviate this tension \cite{Kreisch:2019yzn}\footnote{It was also noted in Ref.~\cite{Kreisch:2019yzn} that including the CMB polarization data in the fit tends to reduce the statistical significance of this scenario, though an earlier study~\cite{Lancaster:2017ksf} found that including the polarization data increases the statistical significance.}. 
The preferred value of the interaction strength is in the ballpark of  $10^{7}\sim10^{9}$ in units of Standard Model (SM) weak interaction strength $G_F$. In this regime neutrino free-streaming is suppressed at high red-shift and  it is not surprising that  such an interaction can have remarkable consequences for the physics of the early Universe.

Large $\nu$SI present a challenge from a particle physics perspective and it is expected that terrestrial experiments can help scrutinize this option. In \cite{Blinov:2019gcj}, the authors explored different options for 
enhanced neutrino interaction. While they found that the vector forces of the aforementioned strength are already disfavored from laboratory experiments, light (below ${\cal O}(10^{2})$ MeV) bosons strongly coupled to neutrinos remain viable. 
The only surviving option which alleviates the Hubble tension is $\nu_\tau$-philic light scalar; this is expected since it is well known that new interactions of $\nu_{\tau}$ are generically the least constrained compared to other flavors. 
Let us note that the authors of \cite{Lyu:2020lps} have recently reached similar conclusion by performing an analysis in the framework of effective theory which respects SM gauge invariance.
It is therefore timely to revisit the constraints on such interactions from particle physics processes and pay particular attention to  the interactions of $\tau$ neutrino.

There are numerous studies of $\nu$SI through the exchange of ``light" mediators in the literature. This class of new physics was explored in meson decays~\cite{Barger:1981vd,Lessa:2007up,Pasquini:2015fjv,Berryman:2018ogk,deGouvea:2019qaz}, double beta decay~\cite{Burgess:1992dt,Burgess:1993xh,Gando:2012pj,Agostini:2015nwa,Blum:2018ljv,Cepedello:2018zvr,Brune:2018sab,Berryman:2018ogk},
invisible $Z$ decays~\cite{Bilenky:1992xn,Machado:2015sha,Berryman:2018ogk} and $\tau$ decays~\cite{Lessa:2007up}.
In addition, it has been recently pointed out that strong $\nu$SI can also play a relevant role in producing sterile neutrino dark matter~\cite{deGouvea:2019phk,Kelly:2020pcy} as well as testing ultralight dark matter scenarios \cite{Krnjaic:2017zlz,Brdar:2017kbt}. 
We would also like to point out further studies involving cosmological~\cite{Boehm:2012gr,Kamada:2015era,Huang:2017egl,Berryman:2018ogk} as 
well as astrophysical (primarily Supernovae) ~\cite{Choi:1987sd,Choi:1989hi,Kachelriess:2000qc,Hannestad:2002ff,Farzan:2002wx,Ng:2014pca,Farzan:2018gtr,Bustamante:2020mep} probes.

For a light scalar interacting with $\nu_{\tau}$ many of the most sensitive probes of new physics connected to $\nu_e$ and $\nu_\mu$ are not sensitive and the two most relevant laboratory bounds arise from $\tau$ and $Z$ decays.
The authors of Ref.~\cite{Lessa:2007up} were the first to estimate the bound on the neutrino coupling to light scalar by studying the former process. 
The reported limit only applies to a particular choice of $m_\phi$ and cannot be extrapolated to the mass range of interest easily. One of our goals in this paper is to derive   
this limit as a function of scalar mass by using state of the art numerical tools. 

In Ref.~\cite{Berryman:2018ogk} the authors present a comprehensive analysis of constraints on light neutrinophilic scalars. What is very interesting for us, in light of couplings to $\nu_\tau$, is the constraint arising from invisible $Z$ decay, namely the process $Z\rightarrow\nu\nu\phi$, where $\phi$ is light scalar. In addition to this process, we will also consider $Z$ invisible decay $(Z\to \bar{\nu}\nu)$ via a triangle loop diagram. Naively such a contribution may appear subdominant since it contains two powers of scalar coupling to neutrinos already at the amplitude level. However, it interferes with the SM tree level diagram and, therefore, the leading contribution is of the same order in the new physics coupling as the $\phi$-bremsstrahlung and should be expected to give a competitive constraint.

The paper is organized as follows. In \cref{sec:Zdecay} 
we present the main results of our investigation of the new physics contribution to invisible $Z$ decays while relegating the details of the calculation to appendices. In \cref{sec:tau} we discuss the procedure for obtaining limits on new physics from leptonic $\tau$-decays.  We analyze the implications of our results for the allowed interaction strength of a neutrino-philic light mediator and comment on the implication for the proposed solution of the Hubble tension in \cref{sec:limit_summary}.   While the motivation for our study is mostly connected to $\tau$ neutrino flavor, for completeness we also present limits for a $\nu_e$ and $\nu_\mu$-philic scalar as well as flavor universal coupling scenario. In \cref{sec:summary} we summarize our results and present our conclusions.

\section{$Z$ decay}
\label{sec:Zdecay}
\noindent
The new neutrino interactions to be considered in this work are parameterized by  
\begin{align}
{\mathcal L}\supset\sum_{\alpha,\thinspace\beta}\frac{1}{2}y_{\alpha\beta}\,\overline{\nu_{\alpha}^{c}}\,P_{L}\,\nu_{\beta}\,\phi+{\rm h.c.}\,,
\label{eq:z-3}
\end{align}
where $\nu_{\alpha}$ and $\nu_{\beta}$ are Dirac spinors
of neutrinos ($\nu_{\alpha}^{c}$ is the charge conjugate of $\nu_{\alpha}$) and $\alpha$ and $\beta$ stand for flavor indices. Furthermore, $y_{\alpha\beta}=y_{\beta\alpha}$
is a symmetric Yukawa matrix and $\phi$ is a scalar field.
Finally, note that the left projector
$P_{L}=(1-\gamma^{5})/2$ ensures that only left-handed neutrinos are involved in the interaction. This interaction term can be generated for instance in the seesaw scenario; the coupling of singlet $\phi$ with right-handed neutrinos induces interaction of $\phi$ with active neutrino states through lepton mixing \cite{Krnjaic:2017zlz}.

\begin{figure}[t]
\centering

\includegraphics[width=0.3\textwidth]{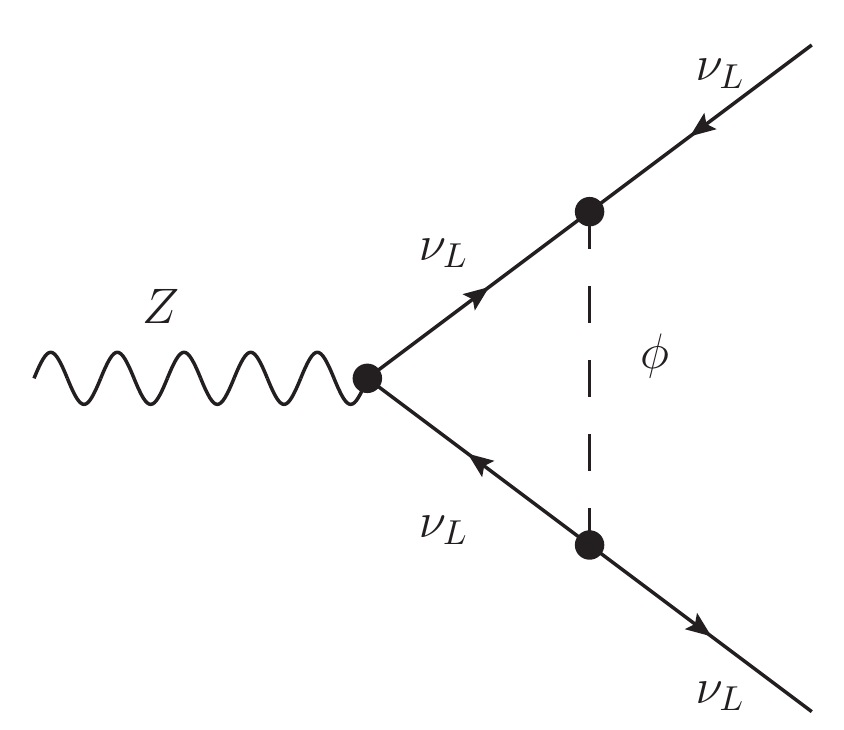}\includegraphics[width=0.3\textwidth]{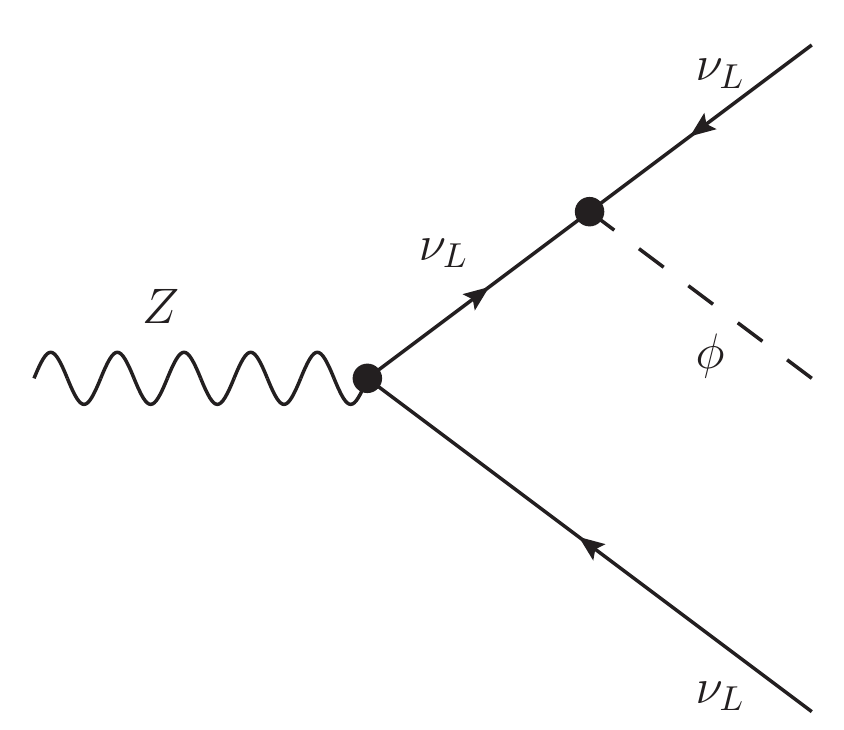}

\caption{Representative Feynman diagrams contributing to invisible $Z$  decays.
\label{fig:Z_decay_feyn}}
\end{figure}

In the presence of these interactions two new physics processes contribute to invisible $Z$ decays and we show the corresponding Feynman diagrams in Fig.~\ref{fig:Z_decay_feyn}. The loop contribution contains two Yukawa vertices being proportional to $|y_{\alpha\beta}|^{2}$ while the bremsstrahlung diagram is proportional to $y_{\alpha\beta}$. Therefore, at the amplitude level, the left diagram is suppressed with respect to the right one by a higher power of Yukawa coupling. However, the loop diagram can interfere with the SM invisible $Z$  decay and this  substantially enhances its contribution to the decay width. Consequently, there is no obvious hierarchy between the two processes and both contributions to the invisible width should be considered in a complete analysis.

\subsection{Loop Contribution}
\noindent
 Let us first consider that two neutrino species with flavors denoted by $\alpha$ and $\beta$ ($\alpha\neq\beta$) are coupled to $\phi$ and other couplings in Eq.~(\ref{eq:z-3}) are absent. The result obtained for this simple case can be easily generalized to the most general Yukawa matrix.

In the presence of a $\overline{\nu_{\alpha}^{c}}P_{L}\nu_{\beta}\phi$ interaction with $\alpha\neq\beta$, there are new physics contributions to both
$Z\rightarrow\bar{\nu}_{\alpha}\nu_{\alpha}$ and $Z\rightarrow\bar{\nu}_{\beta}\nu_{\beta}$ which have identical amplitude and therefore it is enough to only consider $Z\rightarrow\bar{\nu}_{\alpha}\nu_{\alpha}$.
The 1-loop amplitude for $Z\rightarrow\bar{\nu}_{\alpha}\nu_{\alpha}$ reads 
\begin{align}
i{\cal M}(Z\rightarrow\overline{\nu}_{\alpha}\nu_{\alpha})=\frac{i|y_{\alpha\beta}|^{2}}{16\pi^{2}}\epsilon_{\mu}(q)\overline{u}(p_{2})(g_{Z}\gamma^{\mu}P_{L})v(p_{1})\left[\frac{1}{2\epsilon'}+\log\frac{m_{\phi}}{m_{Z}}+\frac{1+i\pi}{2}+{\cal O}\left(\frac{m_{\phi}^{2}}{m_{Z}^{2}}\right)\right]\,,\label{eq:z}
\end{align}
where $g_{Z}$ is the gauge coupling of $Z$ boson to neutrinos; $\epsilon_{\mu}(q)$,
$\overline{u}(p_{2})$ and $v(p_{1})$ denote the external legs associated to the $Z$ boson, neutrino and antineutrino, respectively while $m_{\phi}$ ($m_{Z}$) is the mass of $\phi$ ($Z$). 
The 1-loop diagram for this process is UV divergent. We have adopted gauge invariant dimensional regularization. The typical terms appearing in such calculation are abbreviated by $\epsilon'$
\begin{align}
\frac{1}{\epsilon'}\equiv\frac{1}{\epsilon}-\gamma_{E}+\log(4\pi)+\log\frac{\mu^{2}}{m_{\phi}^{2}}\,.\label{eq:a-4}
\end{align}
Here, $\epsilon=(4-d)/2$ with $d$ representing the number of dimensions, while $\gamma_{E}$ is the Euler-Mascheroni constant.

 The interaction in Eq.~(\ref{eq:z-3}) can also generate other loop diagrams corresponding to neutrino self-energy corrections to the $Z\rightarrow\bar{\nu}_{\alpha}\nu_{\alpha}$ process; see e.g. diagrams in Fig.~\ref{fig:toy_feyn_2}. In the conventional renormalization scheme where only amputated diagrams need to be computed, such diagrams are included by computing the correction to the  $Z$-$\nu$-$\nu$ counter term caused by the wave-function renormalization. Without adding such a counter term, one can also compute these diagrams directly and add them to Eq.~(\ref{eq:z}). These two methods are known to be equivalent; see Appendix~\ref{sec:UV_toy_model} for explicit verification.

Adding the neutrino self-energy corrections to Eq.~(\ref{eq:z}), we obtain
\begin{align}
  i{\cal M}(Z\rightarrow\overline{\nu}_{\alpha}\nu_{\alpha})=\frac{i|y_{\alpha\beta}|^{2}}{16\pi^{2}}\epsilon_{\mu}(q)\overline{u}(p_{2})(g_{Z}\gamma^{\mu}P_{L})v(p_{1})\left[\frac{1}{\epsilon'}+\log\frac{m_{\phi}}{m_{Z}}+\frac{3+2i\pi}{4}+{\cal O}\left(\frac{m_{\phi}^{2}}{m_{Z}^{2}}\right)\right]\,.\label{eq:zzz}
\end{align}
Compared to Eq.~(\ref{eq:z}), the $\log({m_{\phi}}/{m_{Z}})$ term is not changed.

In a complete model, the UV-divergence arising from the considered diagram is expected to be canceled by other diagrams (including counter terms). Here ``complete'' means
not only that all operators should be of dimension 4 or lower, but also that gauge invariance has to be respected. 
We would like to stress that the UV cancellation is model-dependent and, consequently, the finite part can not be predicted fully without committing to a specific UV-completion. Nevertheless, 
the $\log({m_{\phi}}/{m_{Z}})$ term is a generic feature and is independent of the regularization scheme. 
This can for instance be seen by considering only the loop integral  with the loop momentum running between the scales of $m_{\phi}$ and $m_{Z}$ which yields 
a result proportional to $\log({m_{\phi}}/{m_{Z}})$.
This implies that this term can be physically interpreted as the contribution of the loop momentum running in the intermediate scale and  being insensitive to
the UV or IR behavior of the underlying complete models. 
We refer the interested reader to appendix 
\ref{sec:UV_toy_model} where we show the cancellation explicitly in a toy model and find the behavior detailed above.\\

In the SM, the tree-level amplitude for $Z\rightarrow\bar{\nu}_{\alpha}\nu_{\alpha}$ 
is
\begin{align}
i{\cal M}_{{\rm SM}}(Z\rightarrow\bar{\nu}_{\alpha}\nu_{\alpha})=-i\epsilon_{\mu}(q)\overline{u}(p_{2})(g_{Z}\gamma^{\mu}P_{L})v(p_{1}),\label{eq:SM-z}
\end{align}
which leads to the decay width\footnote{The relation between neutrino coupling to $Z$ boson and the weak interaction strength $G_F$ reads $g_{Z}^{2}m_{Z}=\sqrt{2}G_{F}m_{Z}^{3}$.}~\cite{Giunti}
\begin{equation}
\Gamma_{{\rm SM}}(Z\rightarrow\bar{\nu}_{\alpha}\nu_{\alpha})=\frac{G_{F}m_{Z}^{3}}{ 12\sqrt{2}\pi}.\label{eq:z-4}
\end{equation}

By comparing Eq.~(\ref{eq:z}) and Eq.~(\ref{eq:SM-z}), we can obtain
the decay width including the loop contribution which yields
\begin{align}
\Gamma_{{\rm new}}(Z\rightarrow\bar{\nu}_{\alpha}\nu_{\alpha})=\Gamma_{{\rm SM}}(Z\rightarrow\bar{\nu}_{\alpha}\nu_{\alpha})\left|1+\frac{|y_{\alpha\beta}|^{2}}{16\pi^{2}}(L+i \pi/2)\right|^{2},\ \ {\rm with}\ \ L=\log\frac{m_{Z}}{m_{\phi}}+\frac{3}{4}.\label{eq:z-5}
\end{align}

One can check that 
 the final result for the case $\alpha=\beta$ turns out to be the same as Eq.~(\ref{eq:z-5}) with $\beta\rightarrow\alpha$.
Therefore, in the presence of the most general Yukawa matrix, one
only needs to replace $|y_{\alpha\beta}|^{2}$ with $\sum_{\beta}|y_{\alpha\beta}|^{2}$
in Eq.~(\ref{eq:z-5}). If we sum over $\alpha$ indices and restrict ourselves to
terms proportional to the second power of Yukawa coupling or lower we get 
\begin{align}
\Gamma_{{\rm new}}(Z\rightarrow\overline{\nu}\nu)\equiv\sum_{\alpha}\Gamma_{{\rm new}}
(Z\rightarrow\overline{\nu_{\alpha}}\nu_{\alpha})\approx
\frac{G_{F}m_{Z}^{3}}{ 12\sqrt{2}\pi}\left[3+\frac{{\rm tr}[YY^{\dagger}]}{16\pi^{2}}2L\right],\label{eq:z-6}
\end{align}
where $Y$ is the $3\times 3$ Yukawa matrix with  $y_{\alpha\beta}$ elements. One can also
see that Eq.~(\ref{eq:z-6}) is invariant under
 $\nu\rightarrow U\nu$, $Y\rightarrow UYU^{\dagger}$ basis transformations where $U$
is an arbitrary unitary matrix.

\subsection{Bremsstrahlung}
\noindent
The bremsstrahlung process is depicted by the right diagram in Fig.~\ref{fig:Z_decay_feyn}.
Again, we first consider $Z\rightarrow\nu_{\alpha}\phi\nu_{\beta}$
with $\alpha\neq\beta$.  The
decay width in case of $\alpha\neq\beta$ reads
\begin{align}
\Gamma_{{\rm new}}(Z\rightarrow\nu_{\alpha}\phi\nu_{\beta})=\frac{g_{Z}^{2}|y_{\alpha\beta}|^{2}m_{Z}}{24(2\pi)^{3}}F\,,\label{eq:z-8}
\end{align}
where
\[
F\approx\left(1+3\frac{m_{\phi}^{2}}{m_{Z}^{2}}\right)\log\frac{m_{Z}}{m_{\phi}}-\frac{17}{12}.
\]
For details of the derivation including the expression with the full $m_\phi$ dependence of $\Gamma_{{\rm new}}(Z\rightarrow\nu_{\alpha}\phi\nu_{\beta})$ see Appx.~\ref{sec:Z_brem}.
The expression for the total width of the $\phi$
bremsstrahlung with the most general Yukawa couplings is given by 
\begin{eqnarray}
\Gamma_{{\rm new}}(\phi\ {\rm bremsst.}) & = & \frac{1}{2}\sum_{\alpha\neq\beta}\Gamma_{{\rm new}}(Z\rightarrow\nu_{\alpha}\phi\nu_{\beta})+\sum_{\alpha}\Gamma_{{\rm new}}(Z\rightarrow\nu_{\alpha}\phi\nu_{\alpha})\label{eq:z-7}\\
 & = & \frac{1}{2}\sum_{\alpha\neq\beta}\Gamma_{{\rm new}}(Z\rightarrow\nu_{\alpha}\phi\nu_{\beta})+\frac{1}{2}\sum_{\alpha}\left.\Gamma_{{\rm new}}(Z\rightarrow\nu_{\alpha}\phi\nu_{\beta})\right|_{\beta\rightarrow\alpha},\label{eq:z-7-1}
\end{eqnarray}
where the first $1/2$ factor is due to double counting of $\sum_{\alpha\neq\beta}$,
and the last $1/2$ factor accounts for the phase space of identical
particles. In the last term of Eq.~(\ref{eq:z-7-1}), $\Gamma_{{\rm new}}(Z\rightarrow\nu_{\alpha}\phi\nu_{\beta})$
takes the same expression as Eq.~(\ref{eq:z-8}). Eq.~(\ref{eq:z-7-1})
allows the formulation in a basis-independent form similar to Eq.~(\ref{eq:z-6})

\begin{align}
\Gamma_{{\rm new}}(\phi\ {\rm bremsst.})=\frac{1}{2}\frac{g_{Z}^{2}m_{Z}{\rm tr}[YY^{\dagger}]}{24(2\pi)^{3}}F.\label{eq:z-9}
\end{align}

The bremsstrahlung diagram in Fig.~\ref{fig:Z_decay_feyn} represents $Z\rightarrow\overline{\nu}\phi^{*}\overline{\nu}$ process. By flipping the arrows in the diagram one obtains a similar diagram for $Z\rightarrow\nu\phi\nu$
with identical decay width.

Also note that there is the charge conjugate process $Z\rightarrow\overline{\nu}\phi^{*}\overline{\nu}$,
which has the same decay width as $Z\rightarrow\nu\phi\nu$. Therefore, upon
combining all the bremsstrahlung processes we reach the expression for the total
contribution to invisible $Z$ decay 
\begin{align}
\Gamma_{{\rm new}}(\phi/\phi^{*}\ {\rm bremsst.})=\frac{\sqrt{2}G_{F}m_{Z}^{3}{\rm tr}[YY^{\dagger}]}{24(2\pi)^{3}}F.\label{eq:z-10}
\end{align}
We would like to stress that we  
simulated this three-body decay numerically in \texttt{CalcHEP}~\cite{Belyaev:2012qa} and found excellent agreement with our analytic results.


Note that in the limit of $m_{\phi}\rightarrow0$, the sum of loop and bremsstrahlung contribution is divergent. In some theories such as QED, it is well known that the infrared (IR) divergence in the triangle diagram cancels the IR divergence
in the bremsstrahlung diagram. But here one should not expect such
cancellation due to the chirality-flipping feature of scalar interactions.
The processes $Z\rightarrow\overline{\nu}\nu$ and $Z\rightarrow\nu\nu\phi$,
in the limit of $m_{\phi}\rightarrow0$ and zero momentum of $\phi$,
are still physically distinguishable since $\overline{\nu}$ and $\nu$
are different and the IR divergence is regulated by neutrino masses ($m_{\nu}$).
A careful treatment of the case when $m_{\phi}$ is comparable
or lower than $m_{\nu}$ is beyond the scope of this work. Nonetheless,
our results  are valid in regime $m_{\phi}\gg m_{\nu}>0$ that is considered throughout this work.

\section{tau decay}
\label{sec:tau}

\begin{figure}[t]
\centering

\includegraphics[height=0.3\textwidth]{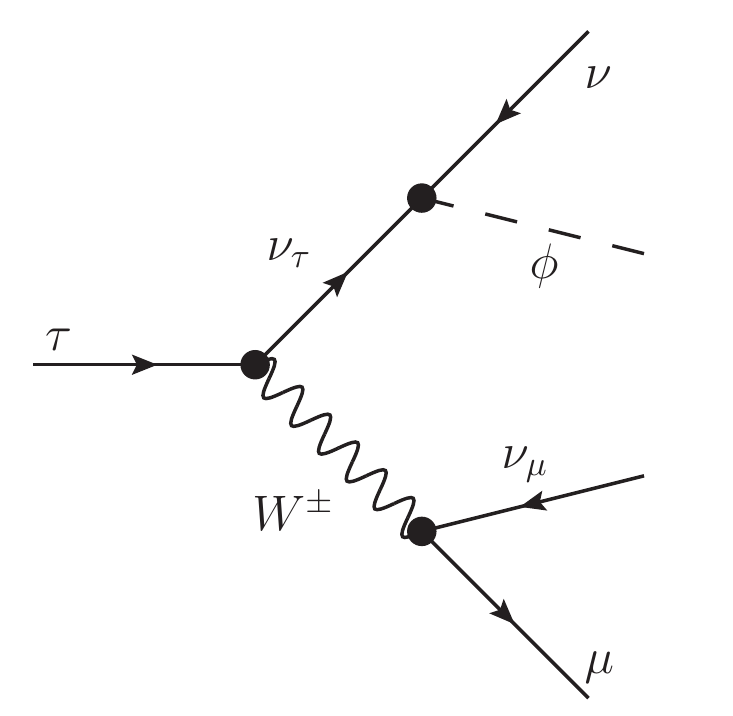}\includegraphics[height=0.3\textwidth]{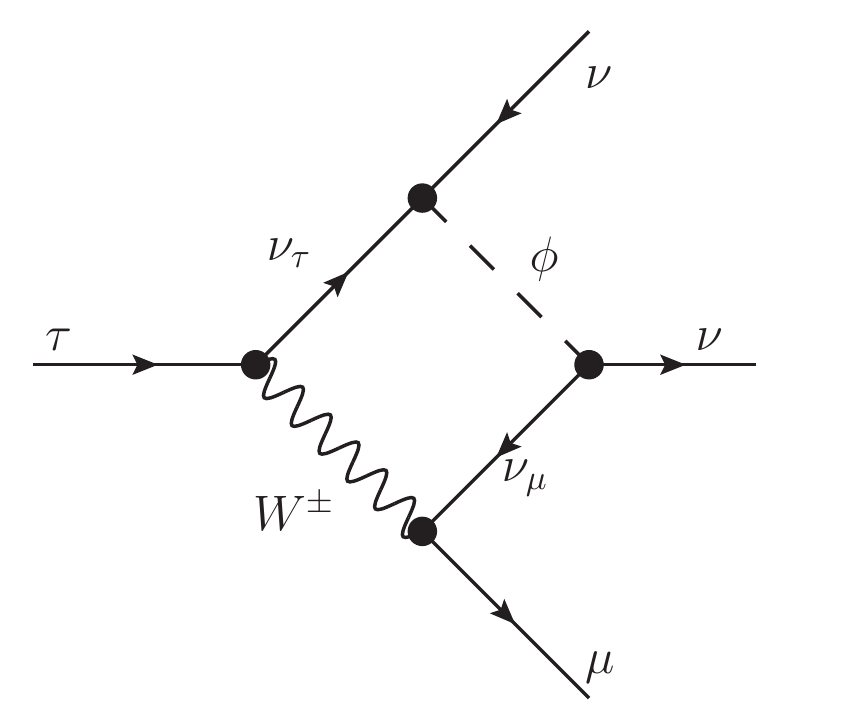}

\caption{Representative Feynman diagrams contributing to $\tau$  decays due to real emission of the new scalar $\phi$ (left panel) and at one loop (right panel).
\label{fig:tau_decay_feyn}}
\end{figure}

\noindent
Another relevant probe of neutrino self-interactions that is particularly relevant for $\nu_\tau$ is the decay of $\tau$ leptons. 
Similar to the $\phi$-bremsstrahlung from Z decays the basic idea here is to constrain the scalar-neutrino coupling  by investigating the impact of attaching  a scalar line to the final state neutrino line; this for instance turns the diagram for the standard three-body decay into a charged lepton (electron or muon) and a pair of neutrinos into a 4-body process containing an extra light scalar boson in the final state. We illustrate this process in the left panel of Fig.~\ref{fig:tau_decay_feyn}. 

Most $\tau$ leptons decay hadronically but with a leptonic branching ratio $Br_{l=e,\mu}\approx 34\%$ the leptonic final states are hardly suppressed. As the leptonic channels are much cleaner we focus on them in the following.
A similar process has been considered previously in the context of a model with light majorons \cite{Lessa:2007up}.
In principle the majoron limits from the literature, available for $m_\phi=1$ keV, could be used to estimate the bounds in the model under consideration here. 
However, as our analysis shows, such bound cannot simply be extrapolated to higher $m_\phi$ and limits derived from rescaling the results of \cite{Lessa:2007up} become unreliable in the mass range of interest here.

\begin{figure}[t!]
  \centering
   \begin{tabular}{ccc}
    \includegraphics[width=0.6\textwidth]{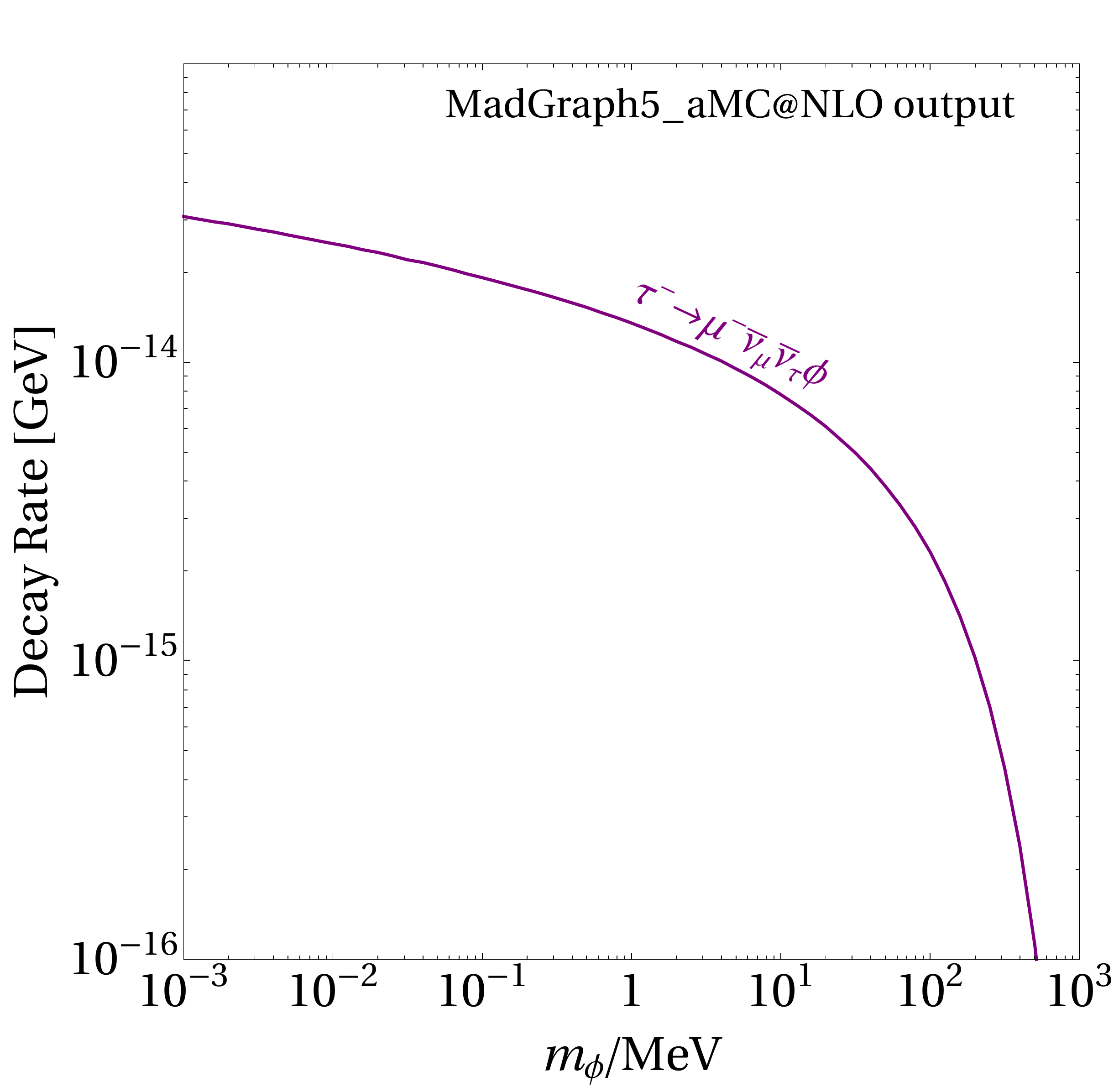}
     \end{tabular}
  \caption{Partial width of the four-body decay $\tau^-\rightarrow\mu^- \bar{\nu}_\mu \bar{\nu}_\tau \phi$ for a representative Yukawa coupling $y_{\tau\tau}=1$. }
  \label{fig:tau}
\end{figure}

We supplement the interaction term in \cref{eq:z-3} to the full SM implementation provided by the \texttt{FeynRules} \cite{Alloul:2013bka} team. Then we generate a \texttt{UFO} model \cite{Degrande:2011ua} which allow us to simulate the process of interest  with \texttt{MadGraph5\_aMC@NLO} \cite{Alwall:2014hca}. As a cross check we first calculate the partial width for $\tau^{-}\to l^{-}\, \bar{\nu}_l \,\nu_\tau$ where $l=\mu^{-}\, \text{or} \,e^{-}$ in the SM and find good agreement with the observed values. We determine the decay rate of the process $\tau^{-}\to l^{-}\, \bar{\nu}_l \,\bar{\nu}_\tau\, \phi$ as a function of $m_\phi$ numerically and construct a fit functions to derive the limit on the Yukawa coupling.
In principle, the rates for the decay into electrons and muons are different due to the different masses of final state particles. In practice,   the discrepancies are expected to be rather small due to the large hierarchy of charged lepton masses. We find that the differences between electron and muon channels are within the numerical uncertainties. The obtained partial width for the Yukawa coupling $y_{\tau\tau}$
equal to $1$ is shown in \cref{fig:tau} as a function of scalar mass, $m_\phi$. This decay rate is used for obtaining the limit as will be demonstrated in \cref{sec:limit_summary}.

Finally, we would like to comment on  another channel that can be constrained from $\tau$ decays. In the diagram for SM process $\tau^{-}\to l^{-}\, \bar{\nu}_l \,\nu_\tau$, the two neutrino lines could be connected with a scalar similar to the loop correction to $Z\rightarrow \nu \bar{\nu}$, see the right panel of Fig.~\ref{fig:tau_decay_feyn} for an illustrative diagram.  Note, however, that in contrast to $Z$ decays only off-diagonal components of the Yukawa lead to a contribution that can interfere with the SM amplitude. 
These off-diagonal couplings are already strongly constrained by meson decays and, therefore, we do not consider this process further. 

\section{Constraints of neutrino interaction with light scalar}
\label{sec:limit_summary}

\begin{figure}[t]
  \centering
  
  \includegraphics[width=0.48\textwidth]{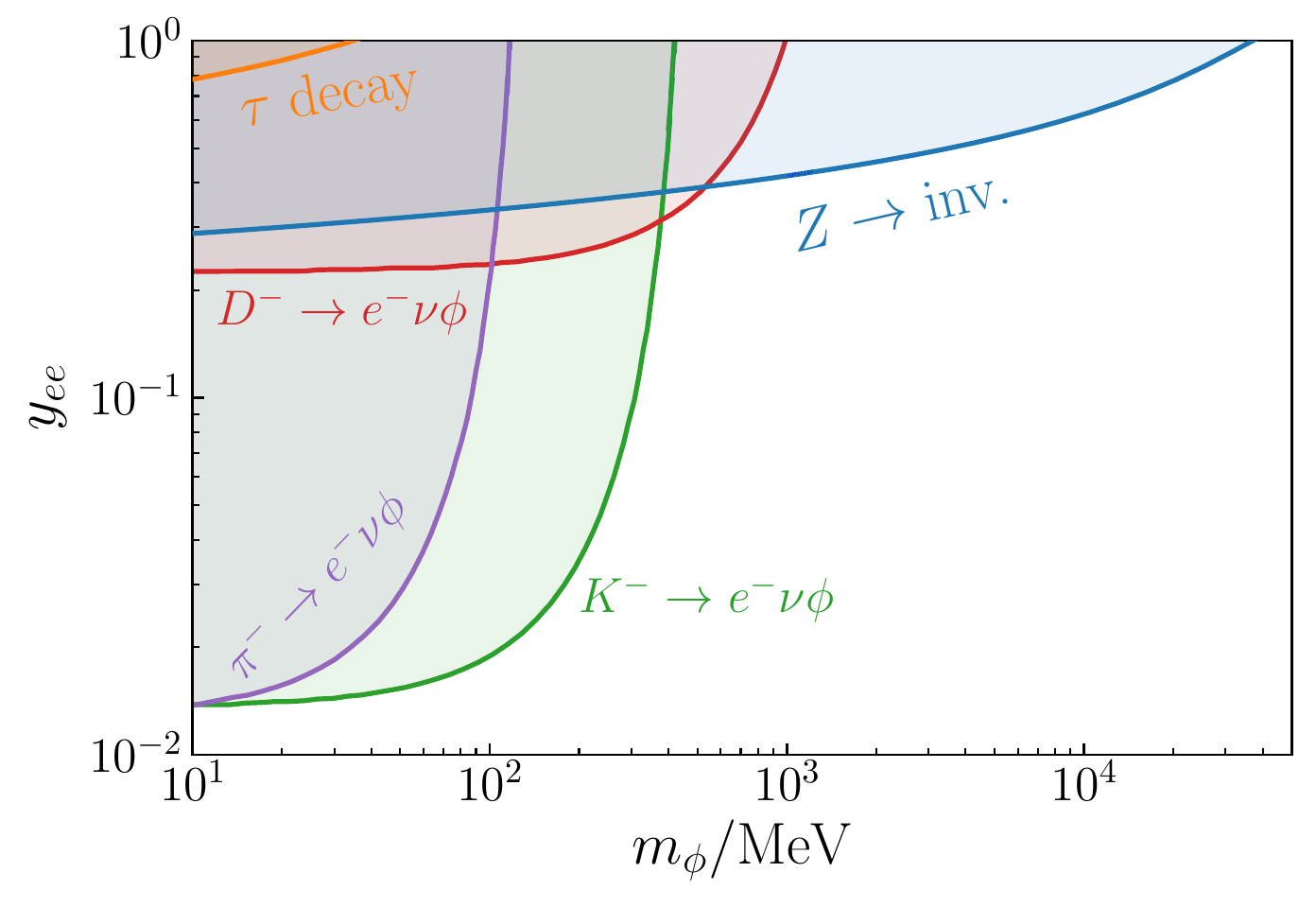}\ \includegraphics[width=0.48\textwidth]{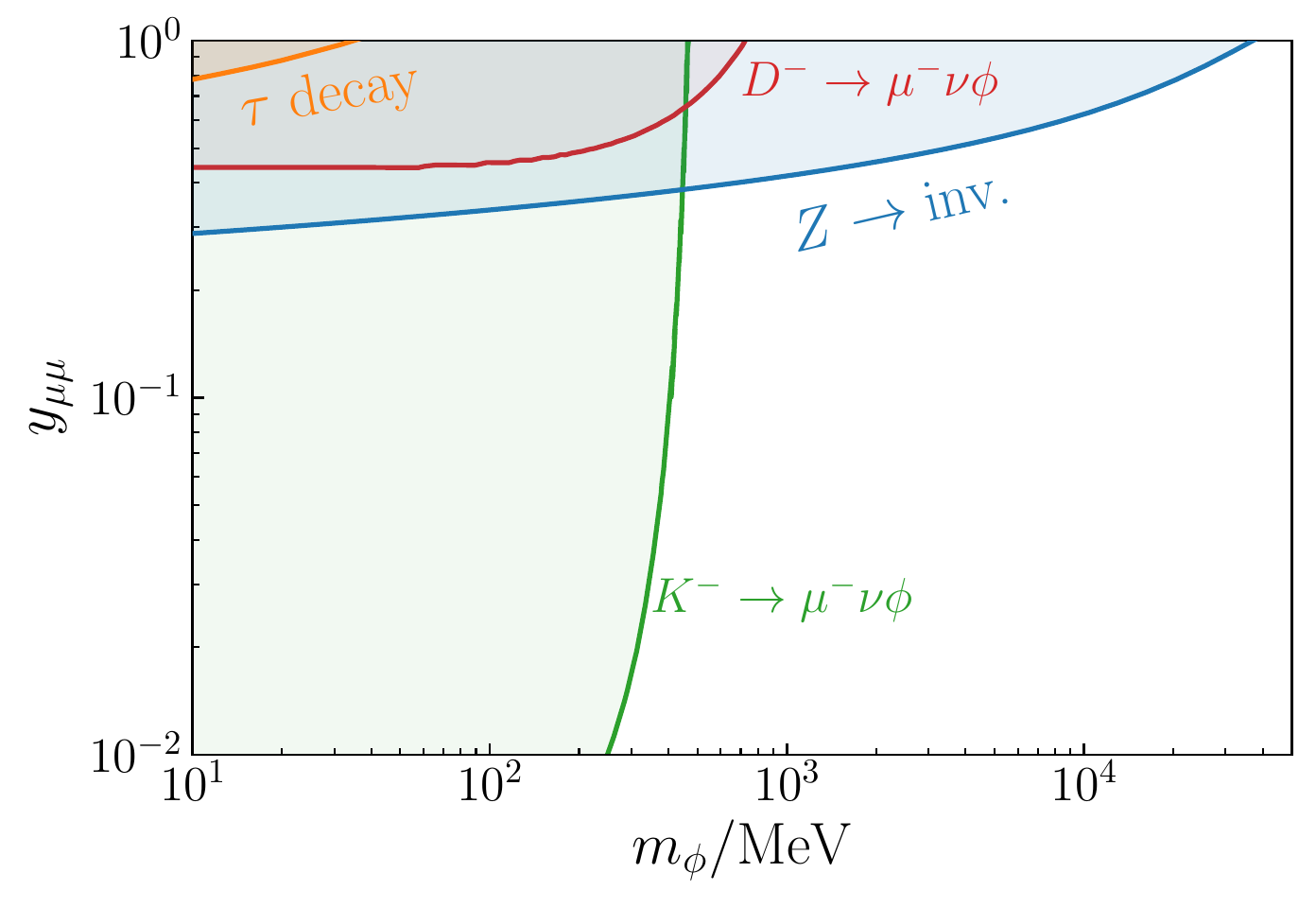}
  
  \includegraphics[width=0.48\textwidth]{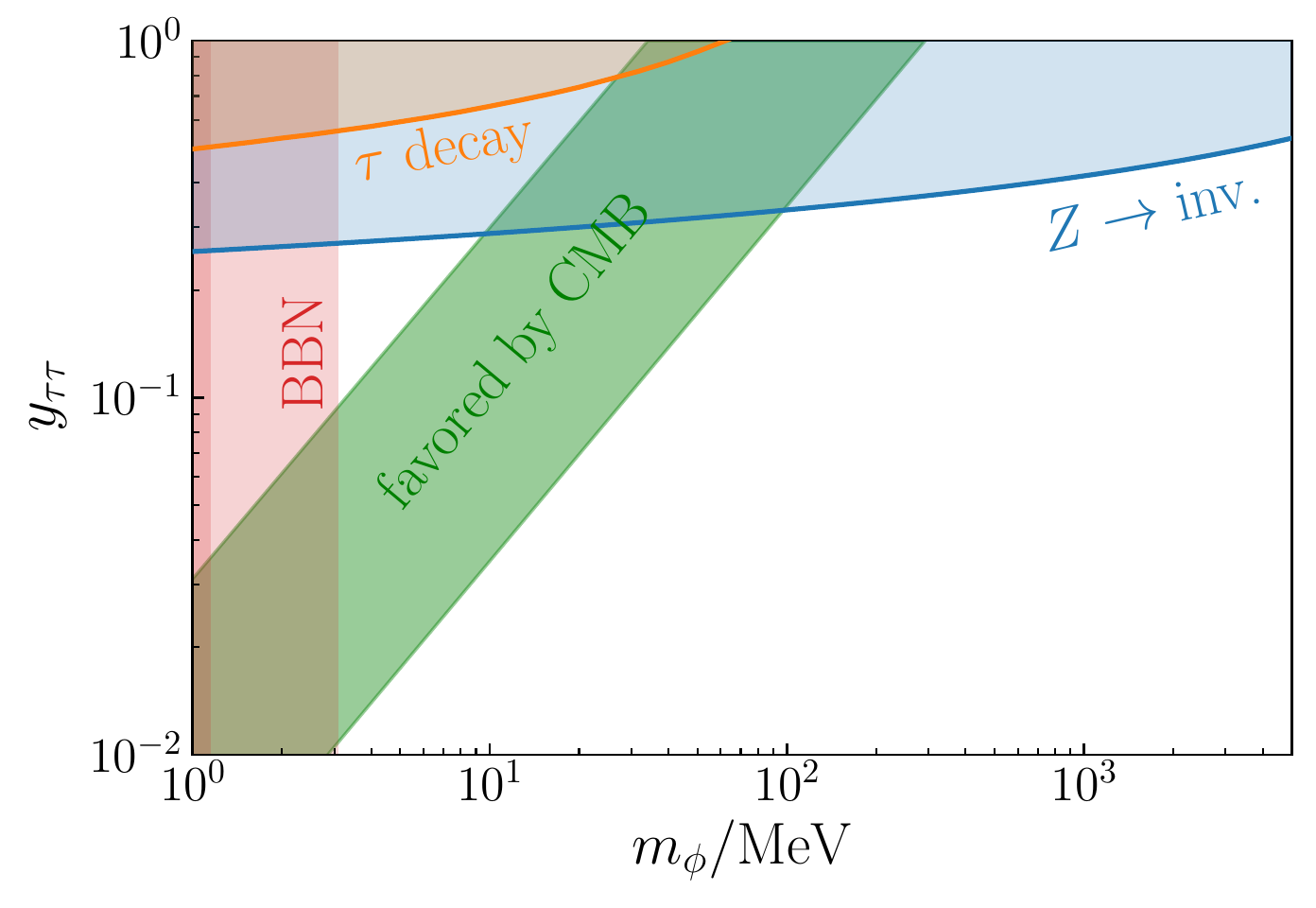}\ \includegraphics[width=0.48\textwidth]{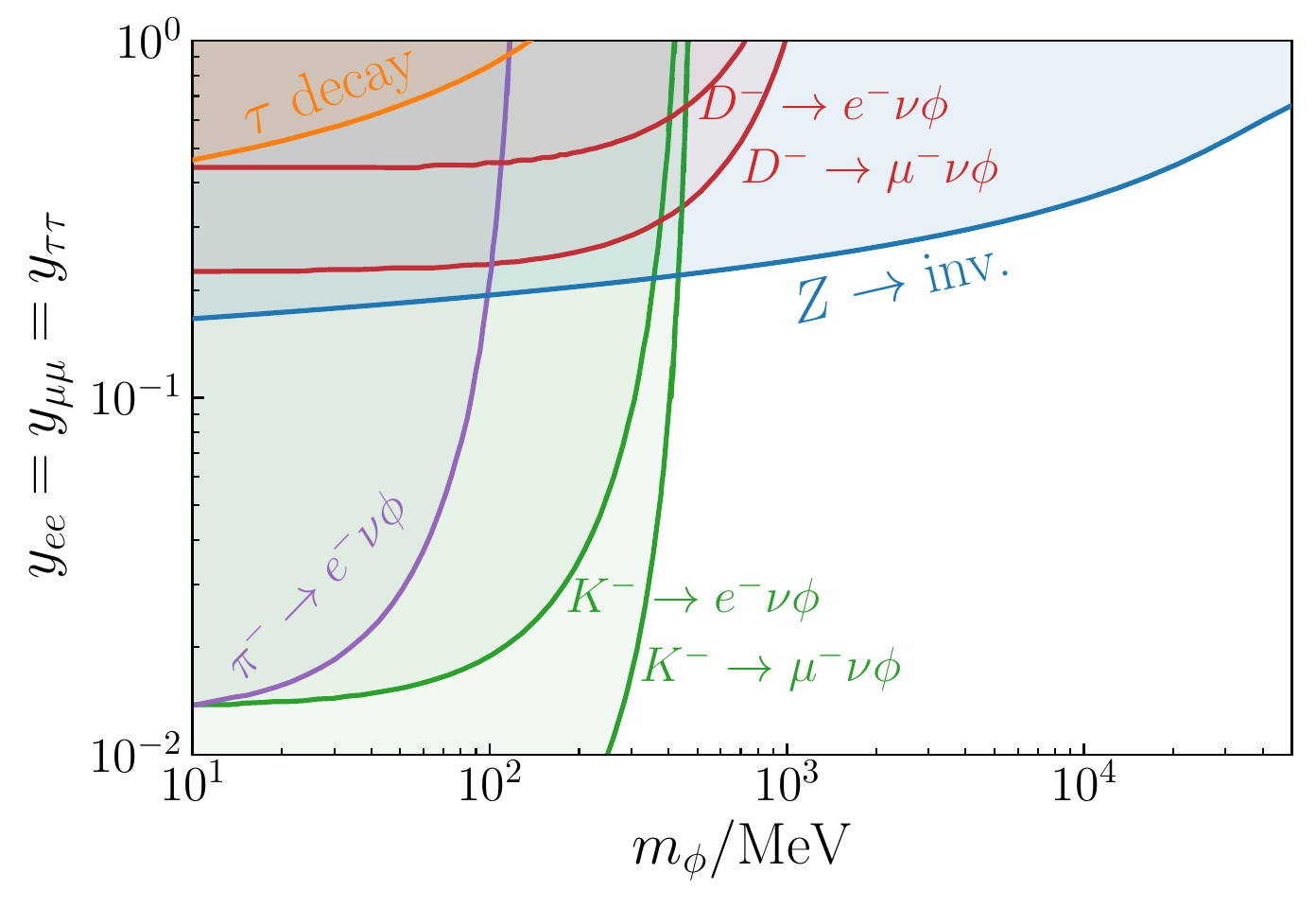}
  
  \caption{Constraints on $\nu$SI from $Z$ invisible decay (blue) and $\tau$ decay (orange) shown together 
  with other known constraints taken from Ref.~\cite{Berryman:2018ogk}. For the case of $\nu_\tau$-philic scalar we also show the preferred region to relax the Hubble tension \cite{Blinov:2019gcj}.  
  \label{fig:Z_decay_bounds}}
  \end{figure}

\noindent
With all the necessary ingredients available we can now turn to actual observables and derive limits on the parameters of the neutrino self-interaction model under consideration here. We will first consider the impact of the measurement of invisible Z decay before turning to $\tau$ decays.

Combining the results in Eqs.~(\ref{eq:z-6}) and (\ref{eq:z-10}),
the total $Z$ invisible width is given by
\begin{align}
\Gamma_{{\rm new}}(Z\rightarrow{\rm {\rm inv.}})\approx
\frac{G_{F}m_{Z}^{3}}{ 12\sqrt{2}\pi}\left[3+\frac{{\rm tr}[YY^{\dagger}]}{16\pi^{2}}2L+
\frac{{\rm tr}[YY^{\dagger}]}{ 8\pi^{2}}F\right].\label{eq:z-11}
\end{align}
Conveniently, the experimental measurement of $Z$ invisible width can be expressed in terms of the number of light neutrino species~\cite{ALEPH:2005ab,Voutsinas:2019hwu,Janot:2019oyi} (see also \cite{Penin:2005kf,Becher:2007cu,Bonciani:2007eh,Actis:2007fs,Actis:2008br}) 
\begin{align}
N_{\nu}=2.9963\pm0.0074,\label{eq:z-12}
\end{align}
which means that the observed invisible width is about 2$\sigma$
lower than the SM prediction. Since both $L$ and $F$ in Eq.~(\ref{eq:z-11}) are positive,
the new physics we introduce can only enhance the $Z$ invisible width.
To get our limits we set the confidence level to $3\sigma$
so that the exclusion bound can be obtained by requiring 
\begin{align}
3+\frac{{\rm tr}[YY^{\dagger}]}{16\pi^{2}}2L+\frac{{\rm tr}[YY^{\dagger}]}{ 8\pi^{2}}F<2.9963+0.0074\times 3.\label{eq:z-13}
\end{align}

In the case of $\tau$ decays the situation is more subtle. Since a $\nu_\tau$ is emitted in every $\tau$ decay a correction to all decay modes is expected for $y_{\tau\alpha}\neq 0$. Naively, one could assume that the correction of the different decay modes is very similar since a $\phi$ emitted from the $\nu_\tau$ is only sensitive to the total momentum of the remaining final state. Consequently, the branching ratios remain similar to the SM prediction while the total width/lifetime of the $\tau$ changes. In contrast, a coupling to $\nu_e$ or $\nu_\mu$ only affects the partial width of the leptonic decay modes. In order to derive reliable bounds on $y_{\tau\tau}$ and $y_{\mu\mu}$  we make use of the partial width $\Gamma_{\tau\mu}$ for the three-body decay $\tau^-\to \mu^-\, \nu_\tau\, \bar{\nu}_\mu$ which can be determined by combining the measured lifetime  $(290.6\pm 1.0) \times 10^{-{15}}$s with the observed branching ratio of $(17.41\pm 0.04)\%$ \cite{Tanabashi:2018oca}.  The central value for the partial decay rate reads $3.94\times 10^{-13}$\,GeV. 
In order to get an estimate of the relative error on the leptonic partial width we add the relative errors of the lifetime and the branching ratio in quadrature and find $\delta \Gamma_{\tau\mu}/\Gamma_{\tau\mu}\approx0.004$.
Therefore, we set the $3\sigma$ exclusion limit on the couplings by requiring 
\begin{align}
\Gamma_{\tau^-\to \mu^-\, \bar{\nu}_\tau\, \bar{\nu}_\mu \phi}\leq3\times 0.004 \times 3.94\times 10^{-13}\,\,\text{GeV}.
\end{align}
A similar procedure utilizing $\tau^-\to e^-\, \bar{\nu}_\tau\, \bar{\nu}_e\phi$ leads to essentially identical results for $y_{ee}$  since $\delta \Gamma_{\tau\mu}/\Gamma_{\tau\mu}\approx 
\delta \Gamma_{\tau e}/\Gamma_{\tau e}$.
When there are contributions to both $\tau^-\to \mu^-\, \bar{\nu}_\tau\, \bar{\nu}_\mu \phi$ 
and $\tau^-\to e^-\, \bar{\nu}_\tau\, \bar{\nu}_e \phi$, we combine both channels together to set our limit.

In Fig.~\ref{fig:Z_decay_bounds}, we present our results; constrains on the diagonal elements of $Y$ are calculated assuming the other elements of the Yukawa matrix are zero. More specifically, for the  case of nonvanishing $y_{ee}$, $y_{\mu\mu}$ and $y_{\tau\tau}$, we take ${\rm tr}[YY^{\dagger}]=|y_{ee}|^{2}$, $|y_{\mu\mu}|^{2}$ and $|y_{\tau\tau}|^{2}$,
respectively. For the $y_{ee}=y_{\mu\mu}=y_{\tau\tau}$ figure (lower right), we take ${\rm tr}[YY^{\dagger}]=|y_{ee}|^{2}+|y_{\mu\mu}|^{2}+|y_{\tau\tau}|^{2}$.
For flavor off-diagonal elements ($y_{\alpha\beta}$ with $\alpha\neq\beta$),
one can simply interpret bounds from any of these
figures as the bounds on $\sqrt{{\rm tr}[YY^{\dagger}]}$ and convert it to the bounds on $y_{\alpha\beta}$. For the coupling of $\nu_e$, $\nu_\mu$ as well as flavor universal scenario (upper panels as well as lower right panel) we also 
superimpose  limits from meson decays \cite{Berryman:2018ogk}. As can be seen these bounds are stronger than those derived in this work for $m_\phi \lesssim 300$ MeV. In the cosmologically most interesting case of a $\nu_\tau$-philic scalar (lower left panel)
we also show the preferred region for alleviating the Hubble tension (green) as well as a constraint from BBN \cite{Blinov:2019gcj}. While the derived laboratory constraints are certainly a relevant player for excluding the parameter space in $y_{\tau\tau}\gtrsim 0.1$ range, 
the viable region still remains in the range $0.1 \gtrsim y_{\tau\tau}\gtrsim 0.01$. This points towards $m_\phi\sim \mathcal{O}(10)$ MeV.

\section{Summary and Conclusions\label{sec:Conclusion}}
\label{sec:summary}
\noindent
In this work we revisited constraints on neutrino self-interactions arising from a neutrino-philic light scalar $\phi$.  The employed probes are invisible $Z$ decays and the leptonic decay modes of the $\tau$.
For invisible $Z$ decays we consider two contributions: one with $\bar{\nu}\nu$ in the final state where we find that 1-loop diagram interfering with the usual SM contribution yields rather significant limit; the other, complementary, contribution to the invisible width arises from bremsstrahlung where two neutrinos (or antineutrinos) appear in the final state alongside $\phi$. Summing both contributions we derive bounds on the new interactions for the case where the light scalar interacts with all flavors individually as well as a flavor universal scenario. 
 In addition, we derive new limit from  leptonic $\tau$ decays. To the best of our knowledge these are the first results that take the dependence on the $\phi$ mass and the coupling fully into account while  previous calculations in the literature only apply for a restricted set of parameters. We
 provide a full picture of our results in \cref{fig:Z_decay_bounds} and compare them to constraints from meson decays. Our results constitute the leading bound on scalars with $m_\phi \gtrsim 300$ MeV irrespective of the preferred flavor. In the case of $\nu_\tau$ self-interactions, which is a particularly relevant scenario in light of recently proposed solution to the Hubble tension, these constraints constitute the leading laboratory limit throughout the considered mass range. However, a scalar in the mass range $10\--100$ MeV remains a viable option for large neutrino self-interactions and we are not able to exclude the whole parameter space preferred by cosmology.
 
\section*{Acknowledgements}
\noindent
The authors would like to thank Ibragim Alikhanov for finding a typo in one of the equations. VB would like to thank Sudip Jana
for discussions regarding usage of Madgraph.

\appendix

\section{Loop calculation in a chiral $U(1)$ toy model \label{sec:UV_toy_model}}
\noindent
In this appendix, we discuss a toy model which is complete and rather minimal containing only a chiral fermion $\nu_{L}$, a gauged $U(1)$ with the gauge
boson denoted as $Z^{\mu}$ and a scalar boson $\phi$. Although
the toy model is not realistic, it illustrates how the UV cancellation
works explicitly and, in addition, shows the potential difference between the results
in an incomplete model with respect to the complete one.

The model is formulated by the following Lagrangian
\begin{align}
{\cal L}\supset\overline{\nu}_{L}i\slashed{D}\nu_{L}+|D_{\mu}\phi|^{2}-m^{2}\phi^{\dagger}\phi-\frac{1}{4}F^{\mu\nu}F_{\mu\nu}-\frac{1}{2}m_{Z}^{2}Z^{\mu}Z_{\mu}-\left[\frac{y}{2}\overline{\nu_{L}^{c}}\phi\nu_{L}+{\rm h.c.}\right].\label{eq:a}
\end{align}

Here, all terms are gauge invariant with the charge assignments $\nu_{L}\sim Q_{\nu}=-1$
and $\phi\sim Q_{\phi}=+2$, except for the gauge boson mass term
$m_{Z}^{2}Z^{\mu}Z_{\mu}$, which can be easily generated by, e.g.,
introducing another scalar that has a charge of $+3$ and a nonzero
VEV. Note that such details are irrelevant for our discussion below. The
covariant derivatives can be explicitly expressed as $D_{\mu}=\partial_{\mu}-igQZ_{\mu}$,
where $Q$ takes $Q_{\nu}$ or $Q_{\phi}$.

\begin{figure}[h]
\centering

\includegraphics[width=0.55\textwidth]{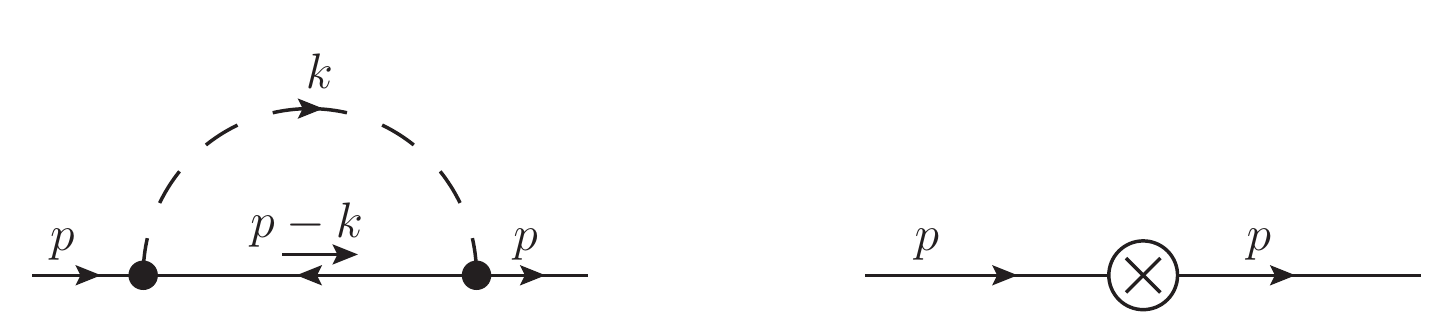}

\includegraphics[width=0.3\textwidth]{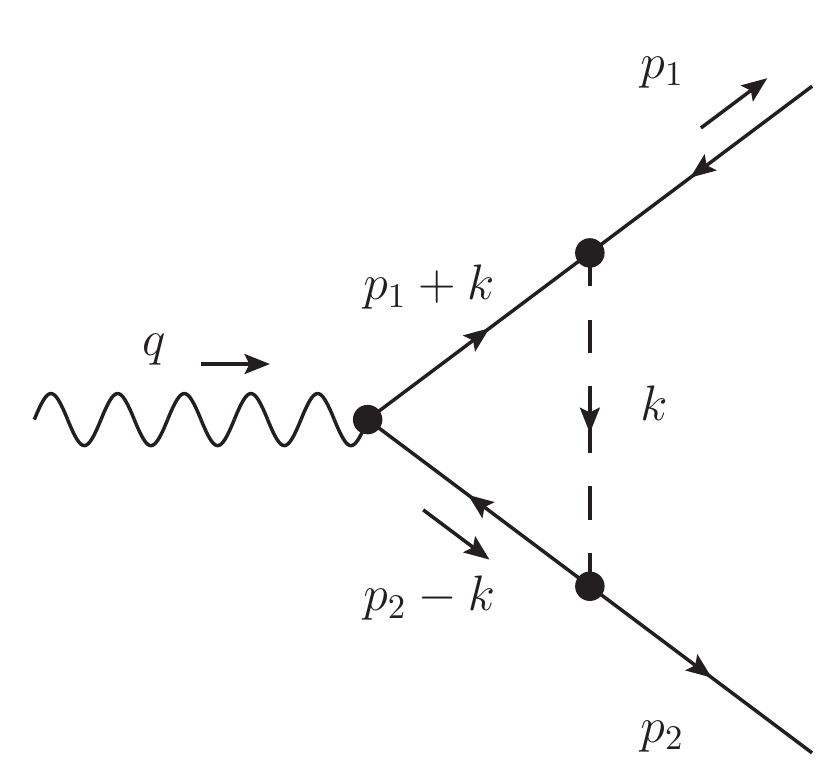}\includegraphics[width=0.3\textwidth]{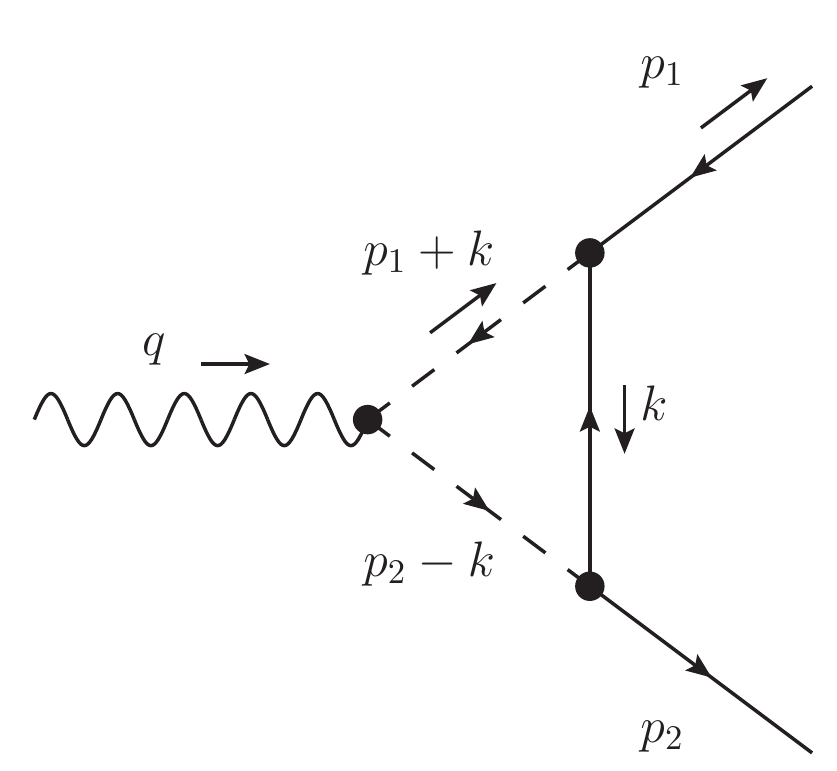}\includegraphics[width=0.3\textwidth]{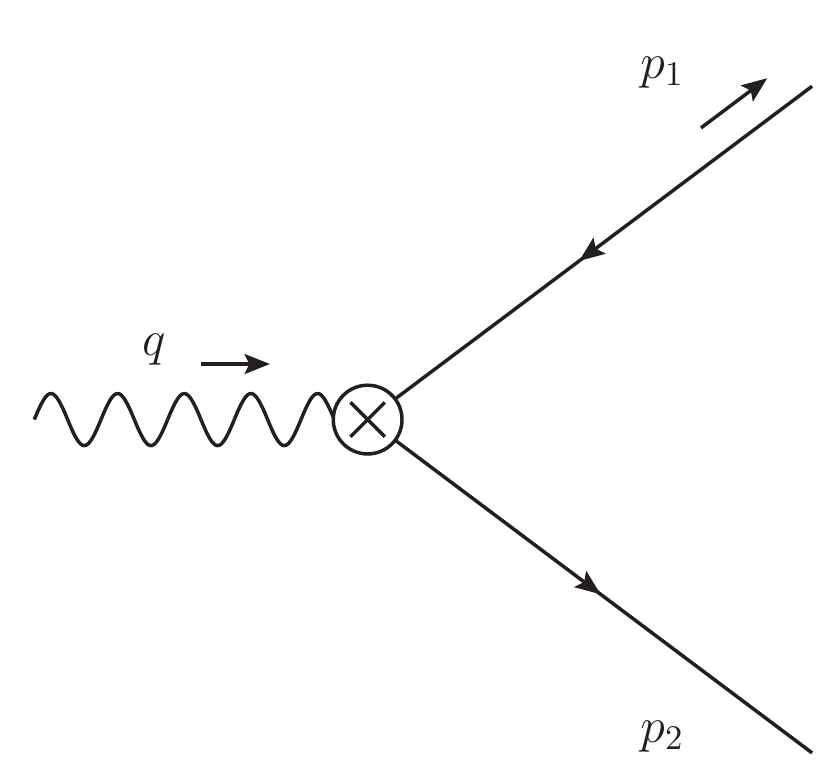}

\caption{Feynman diagrams in the chiral $U(1)$ toy model.\label{fig:toy_feyn}}
\end{figure}

In Fig.~\ref{fig:toy_feyn}, we present the Feynman diagrams involved
in our analyses. We will show explicitly that the UV divergent parts
in these diagrams cancel each other, as long as the $U(1)$ charge
is conserved ($2Q_{\nu}+Q_{\phi}=0$).

First, we compute the 1PI diagram generated by the Yukawa interaction, which
will only lead to renormalization of the wave function of $\nu_{L}$.
It will not lead to mass renormalization as one can expect from the
chiral symmetry, so $\nu_{L}$ remains massless after the loop corrections.
The self-energy generated by the top left diagram in Fig.~\ref{fig:toy_feyn}
reads
\begin{eqnarray}
-i\Sigma(\slashed{p}) & = &  \int 4\frac{d^{4}k}{(2\pi)^{4}}\frac{-iy^{*}}{2}P_{R}\frac{i}{\slashed{p}-\slashed{k}}P_{L}\frac{-iy}{2}\frac{i}{k^{2}-m_{\phi}^{2}}\label{eq:a-1}\\
 & = & |y|^{2}I(p^{2})\slashed{p}P_{L}\,,\label{eq:a-2}
\end{eqnarray}
with 
\begin{align}
I(p^{2})=\frac{i}{16\pi^{2}}\left[\frac{1}{2\epsilon'}+1-\frac{m_{\phi}^{2}}{2p^{2}}+\frac{\left(p^{2}-m_{\phi}^{2}\right){}^{2}}{2p^{4}}\log\frac{m_{\phi}^{2}}{m_{\phi}^{2}-p^{2}}\right].\label{eq:a-3}
\end{align}
Here, we used \texttt{Package-X} \cite{Patel:2015tea} to evaluate
the loop integral. When $p^{2}/m_{\phi}^{2}$ is small, we have the following expansion:
\begin{equation}
  I(p^{2})=\frac{i}{16\pi^{2}}\left[\frac{1}{2\epsilon'}+\frac{1}{4}+\frac{p^{2}}{6m_{\phi}^{2}}+\frac{p^{4}}{24m_{\phi}^{4}}+{\cal O}\left(\frac{p^{6}}{m_{\phi}^{6}}\right)\right].\label{eq:aa-2}
\end{equation}

The UV divergence in the neutrino self-energy is canceled by wave
function renormalization
\begin{equation}
\nu_{L}\rightarrow(1+\delta_{Z})^{1/2}\nu_{L}.\label{eq:a-6}
\end{equation}
The wave function renormalization generates a counter term $\delta_{Z}\overline{\nu_{L}}i\slashed{D}\nu_{L}$
which then can be split to two counter terms $\delta_{Z}\overline{\nu}_{L}i\slashed{\partial}\nu_{L}$
and $\delta_{Z}\overline{\nu}_{L}gQ_{\nu}Z_{\mu}\nu_{L}$. The first
term, corresponding to the top right diagram in Fig.~\ref{fig:toy_feyn},
cancels the UV divergence in Eq.~(\ref{eq:a-3}); while the second
term, corresponding to the bottom right diagram in Fig.~\ref{fig:toy_feyn}
cancels the UV divergences of the two triangle diagrams in Fig.~\ref{fig:toy_feyn}\footnote{Note that in this toy model, if we are only interested in loop corrections
of the Yukawa interactions to the $Z\overline{\nu}_{L}\nu_{L}$ vertex,
then only the wave function renormalization is sufficient to remove
all the UV divergences in the loop diagrams shown in Fig.~\ref{fig:toy_feyn}.}.

Now, by adding the counter term $i\delta_{Z}\slashed{p}P_{L}$ to Eq.~(\ref{eq:a-2})
and requiring the UV cancellation, we obtain
\begin{equation}
\delta_{Z}=i|y|^{2}\left.I(p^{2})\right|_{p^{2}\rightarrow0}=\frac{-|y|^{2}}{16\pi^{2}}\left[\frac{1}{2\epsilon'}+\frac{1}{4}\right].\label{eq:a-5}
\end{equation}

Next, we compute the Feynman diagrams for the $Z_{\mu}\rightarrow\overline{\nu}_{L}\nu_{L}$
decay. The amplitudes of the three bottom diagrams in Fig.~\ref{fig:toy_feyn}
are
\begin{eqnarray}
i{\cal M}_{(a)} & = & \int\frac{d^{4}k}{(2\pi)^{4}}\overline{u}(p_{2})(-iy^{*})P_{R}\frac{i}{\slashed{p}_{2}-\slashed{k}}(-igQ_{\nu^{c}}\gamma^{\mu})\frac{-i}{\slashed{p}_{1}+\slashed{k}}P_{L}(-iy)v(p_{1})\nonumber \\
 &  & \times\frac{i}{k^{2}-m_{\phi}^{2}}\epsilon_{\mu}(q)\,,\label{eq:a-7}\\
i{\cal M}_{(b)} & = & \int\frac{d^{4}k}{(2\pi)^{4}}\overline{u}(p_{2})(-iy^{*})P_{R}\frac{i}{\slashed{k}}P_{L}(-iy)v(p_{1})\nonumber \\
 &  & \times\frac{i}{(p_{2}-k)^{2}-m_{\phi}^{2}}(igQ_{\phi})(p_{2}-p_{1}-2k)^{\mu}\frac{i}{(p_{1}+k)^{2}-m_{\phi}^{2}}\epsilon_{\mu}(q)\,,\label{eq:a-8}\\
i{\cal M}_{(c)} & = & \overline{u}(p_{2})(-ig\delta_{Z}Q_{\nu}\gamma^{\mu}P_{L})v(p_{1})\epsilon_{\mu}(q)\,.\label{eq:a-9}
\end{eqnarray}
Note that $Q_{\nu^{c}}=-Q_{\nu}$ because instead of  $\overline{\nu_{L}}(g\,Q_{\nu})Z_{\mu}\nu_{L}$,
the $Z$-vertex should take the charge conjugate $-\overline{\nu_{L}^{c}}(g\,Q_{\nu})Z_{\mu}\nu_{L}^{c}$
in the left bottom diagram. 
Also note that the neutrino propagators running in the loops are related to $\langle\nu^{c}\overline{\nu^{c}}\rangle$ instead of $\langle\nu\overline{\nu}\rangle$ so when a fermion current arrow is opposite to a momentum arrow in the loops, it implies that the anti-fermion current is aligned with the  momentum arrow.  Hence the numerators above $\slashed{p}_{2}-\slashed{k}$, $\slashed{p}_{1}+\slashed{k}$, and $\slashed{k}$   in   Eqs.~(\ref{eq:a-7}) and (\ref{eq:a-8}) should be $i$, $-i$, and $i$ respectively.
After computing the loop integrals and
expanding the results in $m_{\phi}^{2}/q^{2}$ ($q^{2}=m_{Z}^{2}$),
we obtain 
\begin{equation}
i{\cal M}_{(a)}=\frac{i|y|^{2}Q_{\nu}}{16\pi^{2}}\epsilon_{\mu}(q)\overline{u}(p_{2})(g\gamma^{\mu}P_{L})v(p_{1})\left[\frac{1}{2\epsilon'}+\frac{1}{2}\log\frac{m_{\phi}^{2}}{q^{2}}+\frac{1+i\pi}{2}+{\cal O}\left(\frac{m_{\phi}^{2}}{q^{2}}\right)\right],\label{eq:a-10}
\end{equation}
\begin{equation}
i{\cal M}_{(b)}=\frac{i|y|^{2}Q_{\phi}}{16\pi^{2}}\epsilon_{\mu}(q)\overline{u}(p_{2})(g\gamma^{\mu}P_{L})v(p_{1})\left[\frac{1}{2\epsilon'}+\frac{1}{2}\log\frac{m_{\phi}^{2}}{q^{2}}+\frac{3+i\pi}{2}+{\cal O}\left(\frac{m_{\phi}^{2}}{q^{2}}\right)\right],\label{eq:a-11}
\end{equation}
\begin{equation}
i{\cal M}_{(c)}=\frac{i|y|^{2}Q_{\nu}}{16\pi^{2}}\epsilon_{\mu}(q)\overline{u}(p_{2})(g\gamma^{\mu}P_{L})v(p_{1})\left(\frac{1}{2\epsilon'}+\frac{1}{4}\right).\label{eq:a-12}
\end{equation}
Now we can clearly see that the the UV divergent parts in the above
expressions cancel out if
\begin{equation}
Q_{\nu}+Q_{\phi}+Q_{\nu}=0\;.\label{eq:a-13}
\end{equation}
This corresponds to $Q_{\phi}=-2Q_{\nu}$, which can be understood
from symmetry: $\overline{\nu_{L}^{c}}\phi\nu_{L}$ in Eq.~(\ref{eq:a})
respects the $U(1)$ symmetry only if $Q_{\phi}=-2Q_{\nu}$. 

Taking $Q_{\phi}=-2Q_{\nu}$ and $q^2\rightarrow m_{Z}^{2}$, we get
\begin{equation}
  i{\cal M}_{(a)}+i{\cal M}_{(b)}+i{\cal M}_{(c)}=-\frac{i|y|^{2}Q_{\nu}}{16\pi^{2}}\epsilon_{\mu}(q)\overline{u(p_{2})}(g\gamma^{\mu}P_{L})v(p_{1})\left[\frac{1}{2}\log\frac{m_{\phi}^{2}}{m_{Z}^{2}}+\frac{9+2i\pi}{4}+{\cal O}\left(\frac{m_{\phi}^{2}}{m_{Z}^{2}}\right)\right].\label{eq:aa-10}
\end{equation}

In the above calculation, we have adopted the conventional renormalization scheme which involves counter terms. In such a renormalization scheme, one only needs to compute amputated diagrams while the diagrams in Fig.~\ref{fig:toy_feyn_2} should not be added~\cite{Peskin}. Actually since the counter term $\delta_{Z}\overline{\nu_{L}}i\slashed{D}\nu_{L}$ is generated by the 1PI diagram, the two diagrams in Fig.~\ref{fig:toy_feyn_2} have already been taken into account by the last diagram in Fig.~\ref{fig:toy_feyn}. Nonetheless, it is still interesting to compute the diagrams in Fig.~\ref{fig:toy_feyn_2} to explicitly check that they give result equivalent to the counter term diagram. From Fig.~\ref{fig:toy_feyn_2}, we write down the sum of the two amplitudes:
\begin{figure}[h]
  \centering
  
  \includegraphics[width=0.65\textwidth]{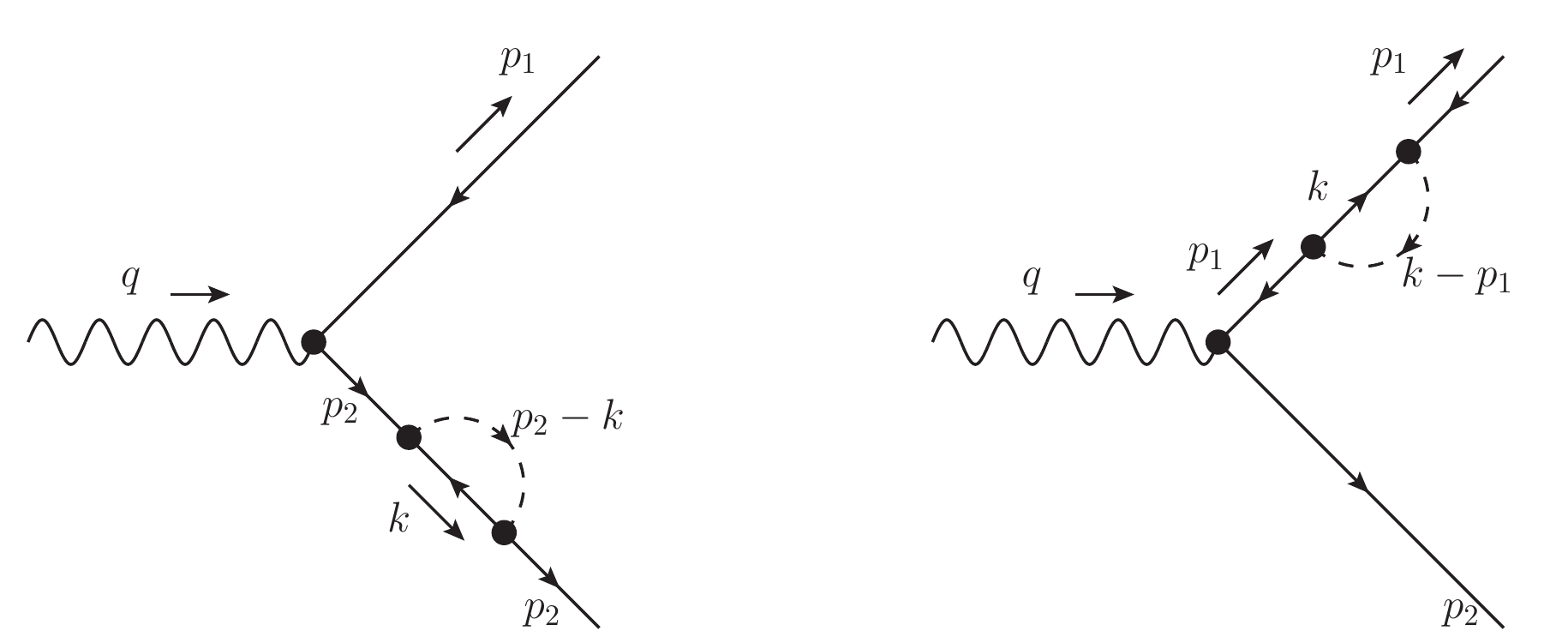}  
  \caption{Loop corrections to the external neutrino legs in the chiral $U(1)$ toy model. These two diagrams are equivalent to the last counter term diagram in Fig.~\ref{fig:toy_feyn}; see the text for more discussions.
  \label{fig:toy_feyn_2}}
\end{figure}
\begin{eqnarray}
  i{\cal M}'_{(c)} & = & \int\frac{d^{4}k}{(2\pi)^{4}}\overline{u(p_{2})}\left[\frac{i}{(p_{2}-k)^{2}-m_{\phi}^{2}}(-iy^{*})P_{R}\frac{i}{\slashed{k}}P_{L}(-iy)\frac{i}{\slashed{p}_{2}-m_{1}}(-igQ_{\nu}\gamma^{\mu})\right.\nonumber \\
   &  & \left.+(-igQ_{\nu}\gamma^{\mu})\frac{-i}{\slashed{p}_{1}-m_{2}}(-iy^{*})P_{R}\frac{-i}{\slashed{k}}P_{L}(-iy)\frac{i}{(k-p_{1})^{2}-m_{\phi}^{2}}\right]v(p_{1})\epsilon_{\mu}(q)\,.\label{eq:aa}
  \end{eqnarray}
  Here we have inserted two masses $m_{1}$ and $m_{2}$ in order to treat singularities properly. At the end of the calculation we will take the zero limit for both. Performing the loop integral, we obtain
  \begin{eqnarray}
  i{\cal M}'_{(c)} & = & \overline{u(p_{2})}|y|^{2}\left[I(p_{2}^{2})\slashed{p}_{2}P_{L}\frac{gQ_{\nu}}{\slashed{p}_{2}-m_{1}}\gamma^{\mu}+\gamma^{\mu}\frac{gQ_{\nu}}{\slashed{p}_{1}-m_{2}}I(p_{1}^{2})\slashed{p}_{1}P_{L}\right]v(p_{1})\epsilon_{\mu}(q)\nonumber \\
   & = & \overline{u(p_{2})}|y|^{2}gQ_{\nu}\left[I(p_{2}^{2})\slashed{p}_{2}\frac{\slashed{p}_{2}\gamma_{L}^{\mu}+m_{1}\gamma_{R}^{\mu}}{p_{2}^{2}-m_{1}^{2}}+\frac{\gamma_{L}^{\mu}\slashed{p}_{1}+\gamma_{R}^{\mu}m_{2}}{p_{1}^{2}-m_{2}^{2}}I(p_{1}^{2})\slashed{p}_{1}\right]v(p_{1})\epsilon_{\mu}(q)\,,\label{eq:aa-1}
  \end{eqnarray}
  where in the second line we have moved $P_{L}$ to the left side of
  $\gamma^{\mu}$ and defined $\gamma_{L/R}^{\mu}\equiv\gamma^{\mu}P_{L/R}$,
  so that all the other gamma matrices can either meet $\overline{u(p_{2})}$
  or $v(p_{1})$. Then using $\overline{u(p_{2})}\slashed{p}_{2}=\overline{u(p_{2})}m_{2}$
  and $\slashed{p}_{1}v(p_{1})=-m_{1}v(p_{1})$, we obtain
  \begin{eqnarray}
  i{\cal M}'_{(c)} & = & \overline{u(p_{2})}|y|^{2}gQ_{\nu}\left[I(m_{2}^{2})\frac{m_{2}^{2}\gamma_{L}^{\mu}+m_{2}m_{1}\gamma_{R}^{\mu}}{m_{2}^{2}-m_{1}^{2}}+\frac{-m_{1}^{2}\gamma_{L}^{\mu}+\gamma_{R}^{\mu}m_{1}m_{2}}{m_{2}^{2}-m_{1}^{2}}I(m_{1}^{2})\right]v(p_{1})\epsilon_{\mu}(q)\nonumber\\
   & \approx & \overline{u(p_{2})}\frac{i|y|^{2}gQ_{\nu}}{16\pi^{2}}\gamma_{L}^{\mu}\left(\frac{1}{2\epsilon'}+\frac{1}{4}+\frac{m_{2}^{4}-m_{1}^{4}}{6m_{\phi}^{2}(m_{2}^{2}-m_{1}^{2})}\right)v(p_{1})\epsilon_{\mu}(q)\,,\label{eq:aa-3}
  \end{eqnarray}
  where in the second line we have used the expansion in Eq.~(\ref{eq:aa-2})
  and ignored higher order terms. In addition, $\gamma_{R}^{\mu}$ terms
  are also ignored because they vanish in the limit of $m_{2}\rightarrow0$
  and $m_{1}\rightarrow0$. Comparing Eq.~(\ref{eq:aa-3}) with Eq.~(\ref{eq:a-12}),
  we can see that $i{\cal M}'_{(c)}=i{\cal M}{}_{(c)}$ in the limit
  of $m_{2}\rightarrow0$ and $m_{1}\rightarrow0$. This verifies that the two diagrams in Fig.~\ref{fig:toy_feyn_2} are indeed equivalent to the counter term diagram in Fig.~\ref{fig:toy_feyn}.

The result in Eq.~(\ref{eq:aa-10}) contains an IR divergence if $m_\phi\rightarrow 0$. In the main text, we have discussed that this result is only valid for $m_Z \gg m_{\phi}\gg m_{\nu}$. 
In the presence of nonzero $m_{\nu}$, one needs to insert $m_{\nu}$ in all the neutrino propagators in Eqs.~(\ref{eq:a-2}), (\ref{eq:a-7}), and (\ref{eq:a-8}). Then following a straightforward but lengthy calculation, we obtain a result which can be written in a form similar to Eq.~(\ref{eq:aa-10}) with the $\frac{1}{2}\log\frac{m_{\phi}^{2}}{m_{Z}^{2}}$ replaced by another function:
\begin{figure}[t]
  \centering    
  \includegraphics[width=0.65\textwidth]{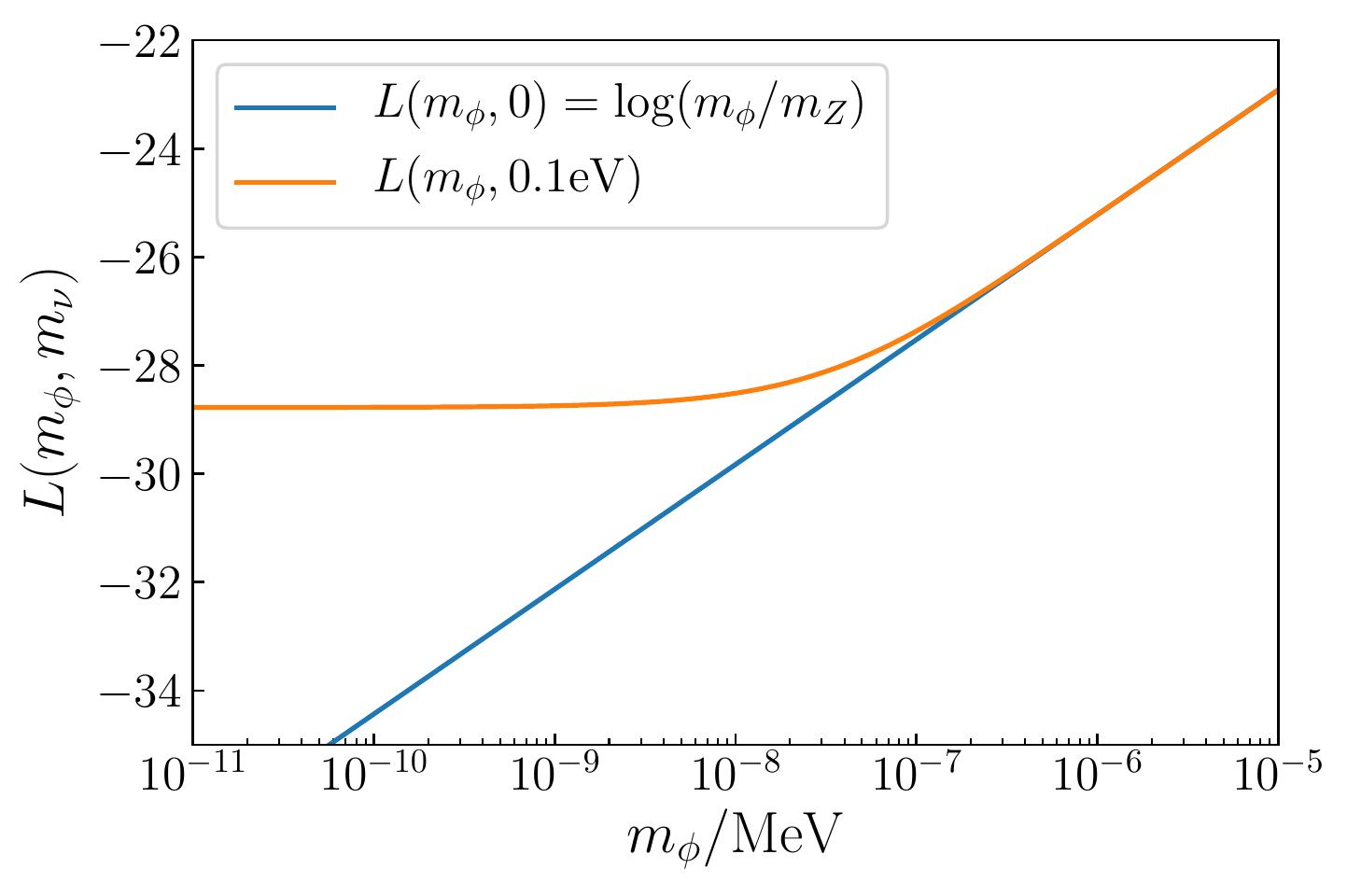}  
  \caption{Numerical evaluation of the $L(m_{\phi},m_{\nu})$ function given in Eq.~(\ref{eq:aa-6}). When $m_{\phi}$ is not well above $m_{\nu}$, the $\frac{1}{2}\log\frac{m_{\phi}^{2}}{m_{Z}^{2}}$ in Eq.~(\ref{eq:aa-10}) should be replaced by $L(m_{\phi},\ m_{\nu})$. The plot shows that the IR divergence in the limit of $m_{\phi}\rightarrow 0$ is removed when $m_{\nu}\neq 0$.  
  \label{fig:log}}
\end{figure}
\begin{equation}
  \frac{1}{2}\log\frac{m_{\phi}^{2}}{m_{Z}^{2}}\rightarrow L(m_{\phi}, m_{\nu}),\label{eq:aa-11}
\end{equation}
where
\begin{eqnarray}
  L(m_{\phi}, m_{\nu}) & = & \frac{1}{4m_{\nu}^{4}m_{Z}^{2}\left(m_{Z}^{2}-4m_{\nu}^{2}\right)}\\
   &   & \times \left[-4m_{\nu}^{4}\log\left(\frac{m_{\phi}^{2}}{m_{\nu}^{2}}\right)\left(-m_{\phi}^{2}\left(4m_{\nu}^{2}+m_{Z}^{2}\right)+m_{Z}^{2}\left(m_{Z}^{2}-2m_{\nu}^{2}\right)+m_{\phi}^{4}\right)\right.\nonumber \\
   &  & -4m_{\nu}^{4}m_{\phi}^{2}\left(-8m_{\nu}^{4}+m_{\phi}^{2}\left(2m_{\nu}^{2}-m_{Z}^{2}\right)+2m_{\nu}^{2}m_{Z}^{2}\right)C_{0}^{\nu\phi\nu}\nonumber \\
   &  & +8m_{\nu}^{4}\left(m_{\phi}^{4}\left(m_{Z}^{2}-2m_{\nu}^{2}\right)+m_{\phi}^{2}\left(8m_{\nu}^{4}-4m_{\nu}^{2}m_{Z}^{2}\right)+m_{\nu}^{2}m_{Z}^{2}\left(m_{Z}^{2}-2m_{\nu}^{2}\right)\right)C_{0}^{\phi\nu\phi}\nonumber \\
   &  & +8m_{\nu}^{6}\left(m_{\phi}^{2}-4m_{\nu}^{2}\right)\Lambda\left(m_{\nu}^{2},m_{\phi},m_{\nu}\right)\nonumber \\
   &  & +2m_{\nu}^{4}\left(8m_{\nu}^{4}+m_{\phi}^{2}\left(2m_{Z}^{2}-4m_{\nu}^{2}\right)+2m_{\nu}^{2}m_{Z}^{2}-m_{Z}^{4}\right)\Lambda\left(m_{Z}^{2},m_{\nu},m_{\nu}\right)\nonumber \\
   &  & -4m_{\nu}^{4}\left(2m_{\nu}^{2}-m_{Z}^{2}\right)\left(m_{Z}^{2}-2m_{\phi}^{2}\right)\Lambda\left(m_{Z}^{2},m_{\phi},m_{\phi}\right)\nonumber \\
   &  & -2m_{\nu}^{2}m_{Z}^{2}\left(m_{\phi}^{2}-2m_{\nu}^{2}\right)\left(4m_{\nu}^{2}-m_{Z}^{2}\right)\Lambda\left(m_{\nu}^{2},m_{\phi},m_{\nu}\right)\nonumber \\
   &  & +2m_{\nu}^{4}\left(4m_{\nu}^{2}-m_{Z}^{2}\right)\left(4m_{\nu}^{2}-5m_{Z}^{2}+2m_{\phi}^{2}\right)\nonumber \\
   &  & -2m_{\nu}^{2}m_{Z}^{2}\left(m_{\phi}^{2}-3m_{\nu}^{2}\right)\left(4m_{\nu}^{2}-m_{Z}^{2}\right)\nonumber \\
   &  & \left.+m_{Z}^{2}m_{\phi}^{2}\left(m_{\phi}^{2}-4m_{\nu}^{2}\right)\left(4m_{\nu}^{2}-m_{Z}^{2}\right)\log\left(\frac{m_{\phi}^{2}}{m_{\nu}^{2}}\right)\right]-\left(\frac{9}{4}+\frac{i\pi}{2}\right).\label{eq:aa-6}
  \end{eqnarray}
  Here $  C_{0}^{\nu\phi\nu}$ and $C_{0}^{\phi\nu\phi}$ involve two-dimensional integrals that have to be evaluated numerically: 
  \begin{eqnarray}
    C_{0}^{\nu\phi\nu} & \equiv & C_{0}\left(m_{\nu}^{2},m_{\nu}^{2},m_{Z}^{2},m_{\nu},m_{\phi},m_{\nu}\right),\label{eq:aa-5}\\
    C_{0}^{\phi\nu\phi} & \equiv & C_{0}\left(m_{\nu}^{2},m_{\nu}^{2},m_{Z}^{2},m_{\phi},m_{\nu},m_{\phi}\right),
  \end{eqnarray}

\begin{eqnarray}
  C_{0}\left(s_{1},s_{2},s_{3};m_{2},m_{1},m_{0}\right) & \equiv & \lim_{\varepsilon\rightarrow0^{+}}\int_{0}^{1}dy\int_{0}^{1-y}dz\left[s_{1}y^{2}+s_{2}z^{2}+\left(s_{1}+s_{2}-s_{3}\right)yz\right.\nonumber \\
   &  & \left.+y\left(-m_{0}^{2}+m_{1}^{2}-s_{1}\right)+z\left(-m_{0}^{2}+m_{2}^{2}-s_{2}\right)+m_{0}^{2}-i\varepsilon\right].\label{eq:aa-8}
 \end{eqnarray}

  The $\Lambda$ functions is defined as
  \begin{equation}
  \Lambda\left(x,y,z\right)\equiv\frac{\log\left[\left(\sqrt{x^{2}-2xy^{2}-2xz^{2}+y^{4}-2y^{2}z^{2}+z^{4}}-x+y^{2}+z^{2}\right)/\left(2yz\right)\right]}{x\left(x^{2}-2xy^{2}-2xz^{2}+y^{4}-2y^{2}z^{2}+z^{4}\right)^{-1/2}}.\label{eq:aa-9}
  \end{equation}

In \cref{fig:log} we show result of numerical evaluation of 
$L(m_{\phi},m_{\nu})$. In particular, it is demonstrated that 
 the IR divergence in the limit of $m_{\phi}\rightarrow 0$ is removed when $m_{\nu}\neq 0$. It is also shown that for $m_\phi \gg m_\nu$, $L(m_{\phi},m_{\nu})$ expectedly converges to
 $\frac{1}{2}\log\frac{m_{\phi}^{2}}{m_{Z}^{2}}$.

\section{Analytical Calculation of three-body invisible $Z$ decay\label{sec:Z_brem}}
\noindent
The amplitude for the process $Z(q)\to \nu_\alpha(p_1) \nu_\beta(p_2) \phi(k)$ reads 
\begin{align}
\mathcal{M}=i \epsilon^*(q) \,g_Z\, y_{\alpha\beta\,}\bar{u}(p_2)\,\left[\frac{\gamma^\mu P_L (\slashed{p_1}+\slashed{k})}{(p_1+k)^2} + \frac{(\slashed{p_2}+\slashed{k}) \gamma^\mu P_L}{(p_2+k)^2}  \,\right]\,v(p_1)\,,
\end{align} 
which leads to
\begin{align}
|\mathcal{M}|^2&=\frac{1}{3}\sum_{\text{polarizations}} \mathcal{M}\mathcal{M^*} = \nonumber \\  
&=\frac{g_Z^2 |y_{\alpha\beta}|^2}{3} \left(
\frac{2(2E_2-m_Z)(2E_1-m_Z)m_Z (8 E_1 E_2 (E_1+E_2)-12 E_1 E_2 m_Z +m_Z^3)}{m_Z^2 (m_Z-2 E_1)^2 (m_Z-2 E_2)^2} \right.-\nonumber \\ & \left.
\frac{2 m_\phi^2 \, (16 E_1 E_2 (E_1 m_Z +E_2 m_Z - E_1 E_2 -m_Z^2)+ m_Z^4)}{m_Z^2 (m_Z-2 E_1)^2 (m_Z-2 E_2)^2} \right)\,,
\label{eq:bigeq}
\end{align}
where we used the expression for the massive vector polarization sum
\begin{align}
\sum_{\text{polarizations}} \epsilon(q)\,\epsilon^*(q) =\left(-g_{\mu\nu}+\frac{q_\mu\,q_\nu}{m_Z^2}\right)\,,
\end{align}
while $E_1$ and $E_2$ are energies of particles with 4-momenta $p_1$ and $p_2$, respectively.
By employing energy conservation $E_1=m_Z-E_2-E_k$ the square matrix element $|\mathcal{M}|^2$ can be expressed only in terms of 2 energies - one of massive $(E_k)$ and one of effectively massless $(E_2)$ final state particle. This allows for a straightforward evaluation of non-trivial three-body phase space integrals.\\

The differential decay rate reads
\begin{align}
d\Gamma=\int_{\frac{m_Z}{2}-\frac{E_k+\sqrt{E_k^2-m_\phi^2}}{2}}^
{\frac{m_Z}{2}+\frac{-E_k+\sqrt{E_k^2-m_\phi^2}}{2}} \frac{1}{16 m_Z (2\pi)^4} \frac{d^3 |\vec{k}|}{\sqrt{E_k^2-m_\phi^2}E_k} \frac{g_Z^2 |y_{\alpha\beta}|^2}{3} f(E_2,E_k)\,  dE_2,
\end{align}
where $f(E_2,E_k)$ is \cref{eq:bigeq} with the aforementioned substitution for $E_1$. 
The integral $g(E_k)=\int f(E_2,E_k) dE_2$ can be evaluated analytically. We obtain the following result  
\begin{align}
g(E_k)=\frac{2\sqrt{E_k^2-m_\phi^2} (-2 E_k m_Z -3 m_Z^2 +m_\phi^2)}{m_Z^2} + 8 E_k \, \text{ArcCoth}\left[\frac{E_k}{\sqrt{E_k^2-m_\phi^2}}\right]\,.
\end{align}

After inferring  $d^3 |\vec{k}|/\sqrt{E_k^2-m_\phi^2}E_k= 4\pi dE_k$, one obtains the expression for the decay rate
\begin{align}
\Gamma(Z\to\nu_\alpha\nu_\beta\phi)&= \frac{4\pi g_Z^2 |\lambda|_{\alpha\beta}^2}{16\times 3 m_Z (2\pi)^4} \int_{m_\phi}^{(m_\phi^2+m_Z^2)/2m_Z} g(E_k) dE_k \nonumber \\
&= \frac{ g_Z^2 |y_{\alpha\beta}|^2 m_Z}{24 (2\pi)^3}\bigg\{\left(1+3 r^2\right)\log\left(\frac{1}{r}\right)-\frac{17-9r^2-9r^4+r^6}{12} \bigg\} \,,
\end{align}
where $r=m_\phi/m_Z$. 
Notice that in case $\alpha=\beta$ we get an extra $1/2$ factor from the phase space.

\bibliographystyle{JHEP}
\bibliography{refs}

\end{document}